\documentclass{aa}

\usepackage{graphicx}
\usepackage[dvips]{color}
\usepackage{txfonts}
\begin{document}
    \title{A new search for planet transits in \object{NGC~6791}.
      \thanks{Based on observation obtained at the
      Canada-France-Hawaii Telescope (CFHT) which is
      operated by the National Research Council of Canada, the
      Institut National des Sciences de l'Univers of the Centre
      National de la Recherche Scientifique of France, and the
      Univesity of Hawaii and on observations obtained at San Pedro
      M\'artir 2.1\,m telescope (Mexico),
      and Loiano 1.5\,m telescope (Italy).}}

  \author{M. Montalto
          \inst{1},
          G. Piotto
          \inst{1},
          S. Desidera
          \inst{2},
          F. De Marchi
          \inst{1},
          H. Bruntt
          \inst{3,4},
          P.B. Stetson
          \inst{5} \\
          A. Arellano Ferro
           \inst{6},
          Y. Momany
          \inst{1,2}
          R.G. Gratton
          \inst{2},
          E. Poretti
          \inst{7},\\
          A. Aparicio
          \inst{8},
          M. Barbieri
          \inst{2,9},
          R.U. Claudi
          \inst{2},
          F. Grundahl
          \inst{3},
          A. Rosenberg
          \inst{8}.
          }

   \authorrunning{M. Montalto et al.}

   \offprints{M. Montalto,  \\
              \email{marco.montalto@unipd.it} }

   \institute{Dipartimento di Astronomia, Universit\`a di Padova,
              Vicolo dell'Osservatorio 2, I-35122, Padova, Italy
              \and
              INAF -- Osservatorio Astronomico di Padova,
              Vicolo dell' Osservatorio 5, I-35122, Padova, Italy
              \and
              Department of Physics and Astronomy, University of Aarhus,
              Denmark
              \and
              University of Sydney, School of Physics, 2006 NSW, Australia
              \and
              Herzberg Institute of Astrophysics, Victoria, Canada
              \and
              Instituto de Astronom\'ia, Universidad Nacional Aut\'onoma
              de M\'exico
              \and
              INAF -- Osservatorio Astronomico di Brera, Via
              E. Bianchi 46, 23807 Merate (LC), Italy
              \and
              Instituto de Astrofisica de Canarias, 38200 La Laguna,
              Tenerife, Canary Islands, Spain
              \and
              Dipartimento di Fisica, Universit\`a di Padova, Italy
             }

   \date{}

\abstract
{Searching for planets in open clusters allows us to study the effects
of   dynamical   environment   on   planet  formation  and  evolution.}
%
{Considering the  strong  dependence of planet  frequency  on  stellar
metallicity,  we   studied   the   metal   rich   old   open   cluster
\object{NGC~6791} and searched for  close-in  planets using  the
transit technique.}  
%
{A   ten-night   observational   campaign   was performed
using  the   Canada-France-Hawaii  Telescope (3.6m),  the  San   Pedro
M\'artir  telescope  (2.1m),  and   the   Loiano   telescope   (1.5m).  
 To  increase the transit  detection probability we also made use
 of the  Bruntt  et  al.~(2003)  eight-nights observational campaign.
Adequate photometric precision for the detection of planetary transits
was achieved.}
%
{ Should the frequency and properties of close-in planets in
 \object{NGC~6791} be
similar to   those orbiting field stars  of  similar metallicity, then
detailed simulations foresee the presence of 2-3 transiting planets.
 Instead,
 we do not confirm  the transit candidates proposed by Bruntt et al.~(2003).
The   probability that the  null detection   is  simply due to  chance
 coincidence is estimated to  be  $3$\%-$10$\%, depending on  the
metallicity assumed for the cluster.}
%
{Possible  explanations  of    the  null-detection   of  transits
include: (i) a lower frequency of close-in  planets in star clusters;
(ii) a smaller planetary radius for  planets orbiting super metal rich
stars; or (iii) limitations in the basic assumptions.
More extensive photometry with $3$--$4$m class telescopes is required
to allow  conclusive inferences   about the  frequency of  planets  in
\object{NGC~6791}.}


   \maketitle
%

\section{Introduction}
\label{s:introduction}

During  the last decade  more  than 200  extra-solar planets have been
discovered.  However, our knowledge of the  formation and evolution of
planetary   systems   remains    largely  incomplete.     One  crucial
consideration is the     role   played  by environment where planetary
systems may form and evolve.

More   than 10\%  of  the extra-solar   planets so far  discovered are
orbiting stars  that are  members  of  multiple systems   (Desidera \&
Barbieri 2007).  Most  of     these  are   binaries with  fairly large
separations (a few  hundred AU).  However,  in few cases,  the  binary
separation      reaches about  10 AU (Hatzes  et al.~\cite{hatzes03};
Konacki 2005), indicating that   planets   can exist even    
in  the presence of fairly strong dynamical interactions.

Another very interesting  dynamical environment is represented by star
clusters,     where the presence  of nearby stars or proto-stars may
affect  the processes  of  planet formation  and evolution  in several
ways.
 Indeed, close stellar encounters may disperse the proto-planetary disks during
the    fairly   short     (about   10    Myr,    e.g.,  Armitage    et
al.~\cite{armitage03}) epoch of  giant planet formation or disrupt the
planetary system after its formation (Bonnell et al. \cite{bonnell01};
Davies  \&  Sigurdsson  \cite{davies01};  Woolfson  \cite{woolfson04};
Fregeau et al.~\cite{fregeau06}).
Another possible disruptive effect is the strong  UV flux from massive
stars, which causes photo-evaporation of dust  grains and thus prevents
planet      formation   (Armitage     \cite{armitage00};   Adams    et
al.~\cite{adams04}).
These effects  are  expected  to  depend on star   density, being much
stronger for globular  clusters (typical current stellar density $\sim
10^{3}$    stars pc$^{-3}$) than  for   the much sparser open clusters
($\leq 10^{2}$ stars pc$^{-3}$).

The    recent discovery   of a  planet   in  the  tight triple  system
\object{HD$188753$}  (Konacki~\cite{konacki}) adds further interest to
the search for planets in star clusters.
In    fact,  the   small     separation between     the planet    host
\object{HD$188753$A} and the pair \object{HD$188753$BC} (about 6 AU at
periastron) makes it very challenging to understand how the planet may
have been formed (Hatzes \& W\"uchterl \cite{hatzes05}). 
Portegies, Zwart \& McMillan  et al.~(\cite{zwart05}) propose that the
planet   formed in  a wide   triple within an  open  cluster  and that
dynamical evolution  successively modified  the  configuration  of the
system. Without observational confirmation  of the presence of planets
in star clusters, such a scenario is purely speculative.

On the observational side, the search  for planets in star clusters is
a  quite challenging task.  Only the  closest open clusters are within
reach  of   high-precision  radial  velocity  surveys   (the most
successful    planet   search   technique).     However,     the
activity-induced    radial    velocity   jitter   limits significantly
the    detectability  of planets    in    clusters as   young  as  the
\object{Hyades} (Paulson et al.~\cite{paulson}).
Hyades red giants have a smaller activity level,  and the first planet
in an open cluster has been recently announced  by Sato et al. (2007),
around $\epsilon\,$ Tau.

The search for photometric transits appears a more suitable technique:
 indeed  it  is possible to monitor  simultaneously  a large number of  cluster
stars.    Moreover,  the     target    stars    may be much   fainter.
However, the transit technique is mostly sensitive to close-in planets
(orbital periods $\leq$ 5 days).

 Space and ground-based wide-field facilities were  also  used to
search   for  planets in the    globular  clusters \object{47 Tucanae}
and \object{$\omega$ Centauri.
These  studies    (Gilliland et  al.~\cite{gilliland};
Weldrake et al.~\cite{weldrake05}; Weldrake et al. 2006)
     reported not a   single  planet detection.
This seemed to indicate that planetary systems are  at least one order
of magnitude less common in globular clusters than in Solar vicinity.}
The lack  of planets in \object{47  Tuc} and \object{$\omega$ Cen} may be  due 
either to the low metallicity of the clusters
(since planet frequency around solar type stars appears to be a rather
strong function   of the metallicity of the    parent star: Fischer \&
Valenti    \cite{fischer05}; Santos  et   al.~\cite{santos04}), or  to
environmental effects caused by the high stellar density (or both).

One   planet has been identified  in  the globular cluster \object{M4}
(Sigurdsson et al.~\cite{sigurdsson03}), but this is a rather peculiar
case, as the   planet is  in a  circumbinary  orbit  around  a  system
including a pulsar and it may have formed in  a different way from the
planets orbiting solar type stars (Beer et al. \cite{beer04}).

Open  clusters are not as  dense as  globular clusters.  The dynamical
and photo-evaporation effects  should      therefore be   less extreme
than in globular  clusters. Furthermore,  their metallicity (typically
solar)   should, in  principle, be  accompanied    by a higher  planet
frequency.

In  the  past  few  years,   some transit searches  were  specifically
dedicated  to      open    clusters:   see   e.g.   von       Braun et
al.~(\cite{vonbraun05}),  Bramich et al.~(\cite{bramich05}), Street et
al.~(\cite{street}),     Burke et al. (2006),    Aigrain et al. (2006)
and references therein.
However, in a typical open cluster of Solar metallicity with
$\sim 1000$ cluster members, less than one star is expected to show a
planetary transit. 
 This depends on the assumption that the planet frequency in open
clusters    is   similar to   that    seen  for   nearby  field  stars
\footnote{0.75\% of stars  with planets with  period less  than 7 days
(Butler et al.~\cite{butler00}), and  a 10\% geometric probability to
observe a transit.}.
Considering the unavoidable       transits detection loss   due to the
observing   window and  photometric  errors,  it  turns  out that  the
probability  of success of such  efforts is  fairly low unless several
clusters are monitored
\footnote{A planet candidate was recently
reported    by    Mochejska    et    al.~(\cite{mochejska06})       in
\object{NGC~$2158$}, but the radius of the transiting object is larger
than any planet known up to now ($\sim\,1.7\,R_{J}$). The companion is
then most likely a very low mass star.}.

    On   the other hand,   the  planet  frequency might   be    higher  
for open  clusters with  super-solar metallicities.     Indeed,    for 
[Fe/H]   between   $+0.2$     and    $+0.4$   the    planet  frequency
around field stars is 2-6 times larger than at solar metallicity.
 However, only a few clusters have been reported to have
metallicities above [Fe/H]$=+0.2$.
The  most  famous is  \object{NGC~$6791$},  a quite massive cluster
 that is at least  $8$    Gyr  old  (Stetson et
al.~\cite{stetson03};   King  et  al.~\cite{king05}, and  Carraro   et
al.~\cite{carraro06}).  As   estimated   by   different   authors, its
metallicity is likely above [Fe/H]=$+0.2$ (Taylor \cite{taylor01}) and
possibly as high as [Fe/H]=$+0.4$ (Peterson et al.~\cite{peterson}).
The most recent high dispersion spectroscopy studies
confirmed the very high metallicity   of the cluster  ([Fe/H]=$+0.39$,
Carraro   et  al.~\cite{carraro06};    [Fe/H]=$+0.47$,  Gratton     et
al.~\cite{gratton06}).
Its old age implies the absence of significant photometric variability
induced  by  stellar activity.   Furthermore,  \object{NGC~6791} is  a
fairly rich cluster. All these facts make it an almost ideal target.

\object{NGC~6791} has been the target of two photometric campaigns 
 aimed at detecting planets transits.
Mochejska et al.~(\cite{mochejska02},
\cite{mochejska05}, hereafter M05) observed the cluster in the {\em R}
band with the 1.2 m Fred Lawrence Whipple  Telescope during 84 nights,
over three   observing  seasons (2001-2003).   They   found  no planet
candidates, while the expected  number of detections  considering their
photometric precision and observing window was $\sim1.3$.
Bruntt et al.~(\cite{bruntt03},   hereafter B03) observed  the cluster
for 8 nights  using ALFOSC at  NOT.  They found 10 transit candidates,
most of which (7) being likely due to instrumental effects.

Nearly continuous,  multi-site  monitoring lasting several days  could
strongly enhance the transit detectability.
This idea inspired  our  campaigns   for multi-site transit     planet
searches in the super metal rich open clusters \object{NGC~6791} and
\object{NGC~6253}.
This  paper  presents the results of   the observations of the central
field of  \object{NGC~6791},   observed at  CFHT,  San Pedro  M\'artir
(SPM), and Loiano.
 We also made use of the  B03 data-set [obtained at the  Nordic
Optical   Telescope (NOT)  in  $2001$] and  reduce  it  as done for
our  three data-sets.
The  analysis    for    the  external fields,  containing mostly field
stars, and of   the  \object{NGC~6253}  campaign, will be    presented
elsewhere.

The outline of  the paper is the following: Sect.~\ref{s:observations}
presents the instrumental setup and the observations. We then describe
the reduction procedure  in Sect.~\ref{s:reduction}, and the resulting
photometric   precision   for  the   four   different   sites  is   in
Sect.~\ref{s:photometricprecision}.   The selection of cluster members
is     discussed      in     Sect.~\ref{s:selection}.      Then,    in
Sect.~\ref{s:algorithm},  we  describe  the adopted  transit detection
algorithm.    In   Sect.~\ref{s:simul}  we   present  the  simulations
performed to establish the  transit detection efficiency (TDE) and the
false    alarm   rate   (FAR)       of    the algorithm    for     our
data-sets. Sect.~\ref{s:diffapproach}    illustrates    the   different
approaches   that   we   followed  in the    analysis  of    the data.
Sect.~\ref{s:candidates} gives details about the transit candidates.

In Sect.~\ref{s:expected}, we estimate  the expected planet  frequency
around main sequence stars of the cluster, and  the expected number of
detectable    transiting     planets    in     our     data-sets.    In
Sect.~\ref{s:significance}, we compare the results of the observations
with  those of the simulations, and  discussed their significance.  In
Sect.~\ref{s:implications}  we discuss  the different  implications of
our results, and, in  Sect.~\ref{s:comparison},  we make a  comparison
with     other  transit     searches  toward    \object{NGC~6791}.  In
Sect.~\ref{s:future}    we critically   analyze  all the  observations
dedicated  to the search for planets  in \object{NGC~6791} so far, and
propose   future observations    of  the  cluster,  and   finally,  in
Sect.~\ref{s:conclusions}, we summarize our work.

\section{Instrumental setup and observations}
\label{s:observations}

The observations  were acquired during a ten-consecutive-day observing
campaign, from  July 4 to July   13, 2002.  Ideally,      one   should
monitor the cluster nearly continuously.
For this reason,   we used three  telescopes, one  in Hawaii, one   in
Mexico, and the third one in Italy.  In Table~\ref{tab:6791} we show a
brief summary of our observations.

In    Hawaii,   we     used     the    CFHT   with      the     CFH12K
detector~\footnote{www.cfht.hawaii.edu/Instruments/Imaging/CFH12K/},
a mosaic of 12  CCDs of 2048$\times$4096 pixels, for  a total field of
view  of 42$\times$28    arcmin,   and  a  pixel  scale   of   $0.206$
arcsec/pixel.  We acquired  $278$  images of  the cluster  in  the $V$
filter.  The seeing  conditions ranged between
 $0\farcs6$ to $1\farcs9$, with
a  median of $1\farcs0$. Exposure times  were  between $200$ and $900$
sec, with a median value of $600$ sec.
The observers were H.~Bruntt and P.B.~Stetson.

In San Pedro  M\'artir, we used  the 2.1m telescope, equipped with the
Thomson 2k detector.  However the data section of the CCD corresponded
to  a   $1k\,\times\,1k$ pixel array.     The  pixel scale   was  0.35
arcsec/pixel, and therefore the field of  view ($\sim$ 6 arcmin$^{2}$)
contained just  the center of the cluster,   and was smaller  than the
field covered with the other detectors.
We made use of $189$  images taken between July  6, 2002 and July  13,
2002. During the first  two nights the  images were acquired using the
focal reducer,  which increased  crowding and reduced  our photometric
accuracy.  All  the images were taken  in the $V$ filter with exposure
times of $480-1200$ sec (median $660$ sec), and seeing
between $1\farcs1$ and  $2\farcs1$  (median $1\farcs4$).  Observations
were taken by A.~Arellano Ferro.

In Italy, we  used the Loiano 1.5m telescope\footnote{The observations
were originally planned  at the  Asiago Observatory  using the  1.82 m
telescope + AFOSC. However,  a major failure of instrument electronics
made  it impossible to perform  these  observations.  We obtained four
nights  of observations at    the Loiano Observatory,  thanks  to  the
courtesy  of the scheduled observer M.~Bellazzini  and of the Director
of Bologna Observatory F.~Fusi Pecci.}
equipped with BFOSC + the EEV 1300$\times$1348B detector.
The pixel scale was  0.52 arcsec/pixel, for a  total field coverage of
11.5 arcmin$^{2}$.  We observed  the target during four nights ($2002$
July $6-9$). We acquired and reduced $63$ images of the cluster in the
$V$ and Gunn $i$ filters ($61$ in $V$, $2$ in $i$).
The seeing
values were between $1\farcs1$  and  $4\farcs3$ arcsec, with a  median
value  of $1\farcs4$ arcsec.    Exposure times ranged between  120 and
1500 sec (median $1080$ sec).  The observer was S.~Desidera.

We   also   make   use of   the  images taken  by  B03  in 2001.   We
obtained these     images from the   Nordic  Optical   Telescope (NOT)
archive. As explained in BO3, these data  covered eight nights between
July 9 and 17, 2001.
The detector was ALFOSC\footnote{ALFOSC  is owned by the Instituto  de
Astrofisica  de  Andalucia (IAA) and operated   at  the Nordic Optical
Telescope  under agreement between     IAA  and the  NBIfAFG of    the
Astronomical Observatory of Copenhagen.}  a 2k$\times$2k thinned Loral
CCD with a pixel scale of 0.188 arcsec/pixel yielding a total field of
view of 6.5 arcmin$^{2}$. Images were taken in the $I$ and $V$ filters
with  a median seeing of $1$  arcsec. We used only  the  images of the
central part of the cluster (which  were the majority) excluding those
in the external regions. In  total we reduced $227$  images in the $V$
filter and $389$ in the $I$ filter.

It should  be noted that  ALFOSC and  BFOSC are  focal reducers, hence
light concentration
introduced a variable background. These effects  can be important when
summed   with flat fielding   errors.  In  order to reduce       these
un-desired effects,  the images were acquired while trying to maintain
the  stars  in the same   positions  on the  CCDs.   
Nevertheless, the precision  of this pointing procedure was  different
for the four telescopes: the median values
of the telescope shifts are  of $0\farcs5$, $0\farcs5$, $3\farcs4$ and
$2\farcs1$, respectively for Hawaii, NOT,  SPM, and Loiano. This means
that while for    the CFHT and  the NOT   the median shift    was at a
sub-seeing  level (half of the    median seeing),  for the other   two
telescopes it  was respectively of the order  of 2.4 and 1.5 times the
median seeing. Hence,  it  is possible that  flat-fielding  errors and
possible  background variation have  affected the NOT, SPM, and Loiano
photometry, but the  effects on the Hawaii  photometry are expected to
be smaller. 

\begin{table*}[!]
\caption{Summary of the observations taken during
 $4$-$13$ July, $2002$ in Hawaii, San Pedro M\`artir, Loiano
 and from $9$-$17$ July, $2001$ in La Palma (by B03).
\label{tab:6791}
}

\begin{center}
\begin{tabular}{c c c c c}
\hline
                    & Hawaii   &  San Pedro M\'artir &  Loiano & NOT\\
\hline
N. of Images              &  278  &  189  &  63 &   227(V), 389(I)  \\
Nights              &  8   &   8  &  4 & 8  \\
Scale (arcsec/pix)  & 0.21    &   0.35    &  0.52 & 0.188 \\
FOV (arcmin)        & 42 x 28  &   6 x 6 & 11.5 x 11.5 & 6.5 x 6.5 \\
\hline

\end{tabular}
\end{center}
\end{table*}

Bad weather  conditions  and  the  limited time  allocated   at Loiano
Observatory caused incomplete coverage of the scheduled time interval.
Moreover we   did   not  use  the  images  coming from the  last night
(eighth night) of observation in La Palma  with the NOT because of bad
weather conditions.  Our observing window, defined  as the interval of
time    during which  observations  were carried    out,  is shown  in
Tab.~\ref{tab:obswindow} for Hawaii, SPM  and Loiano observations  and
in Tab.~\ref{tab:obswindow1} for La Palma observations.

\begin{table}
\caption{
The observing window relative to the July $2002$ observations. 
The number of observing hours is given for each night.
The last line shows the total number of observing hours for each site.
\label{tab:obswindow}
}
\begin{center}
\begin{tabular}{c c c c}
\hline
 Night & Hawaii &  SPM   &  Loiano \\
\hline
$1^{st}$   & $3.58$   &  $-$  &  $-$   \\
$2^{nd}$   & $3.88$   &  $-$  &  $-$   \\
$3^{rd}$   & $2.68$   &  $6.31$  &  $3.58$   \\
$4^{th}$   & $7.56$   &  $6.61$  &  $5.21$   \\
$5^{th}$   & $5.23$   &  $6.58$  &  $6.08$   \\
$6^{th}$   & $8.30$   &  $6.86$  &  $5.45$   \\
$7^{th}$   & $-$   &  $3.20$  &  $-$   \\
$8^{th}$   & $-$   &  $7.03$  &  $-$   \\
$9^{th}$   & $8.34$   &  $7.27$  &  $-$   \\
$10^{th}$  & $8.37$   &  $4.15$  &  $-$   \\      \hline
Total      &  \bf{$47.94$} & \bf{$48.01$} & \bf{$20.32$}  \\
\hline
\end{tabular}
\end{center}

\caption{
The observing window relative to the July $2001$ observations. 
\label{tab:obswindow1}
}
\begin{center}
\begin{tabular}{c c}
\hline
 Night & La Palma \\
\hline
$1^{st}$   &   $5.45$ \\
$2^{nd}$   &   $7.06$ \\
$3^{rd}$   &   $8.04$ \\
$4^{th}$   &   $7.61$ \\
$5^{th}$   &   $7.57$ \\
$6^{th}$   &   $7.82$ \\
$7^{th}$   &   $7.72$ \\
$8^{th}$   &   $2.41$ \\
\hline
Total      &  \bf{$53.68$}  \\
\hline
\end{tabular}
\end{center}
\end{table}

\section{The reduction process}
\label{s:reduction}

\subsection{The pre-reduction procedure}
For the San Pedro, Loiano and La Palma images, the pre-reduction was 
done in a standard way, 
using  IRAF   routines\footnote{IRAF is distributed  by  the  National
Optical Astronomy Observatory, which is operated by the Association of
Universities for    Research in   Astronomy, Inc.,   under cooperative
agreement with the NSF.}.   The images  from  the Hawaii  came already
reduced                  via                  the               ELIXIR
software\footnote{http://www.cfht.hawaii.edu/Instruments/Elixir}.

\subsection{Reduction strategies}
The data-sets described in Sec.~\ref{s:observations} were reduced with
three different   techniques:   aperture  photometry,   PSF    fitting
photometry  and  image subtraction. An   accurate description of these
techniques is given in  the next Sections.    Our goal was  to compare
their performances to see  if one of them   performed better than  the
others. 
For what concerned aperture  and PSF  fitting  photometry we used  the
DAOPHOT (Stetson 1987) package.  In particular the aperture photometry
routine was slightly  different from that one  commonly used in DAOPHOT
and was provided by P~.B.~Stetson.  It  performed the photometry after
subtracting all the    neighbors  stars of  each target   star.  Image
subtraction was  performed by means of the  ISIS2.2 package  (Alard \&
Lupton, 1998) except for what  concerned  the final photometry on  the
subtracted  images which  was performed    with the DAOPHOT   aperture
routine       for               the      reasons    described       in
paragraph~\ref{s:imagesubtraction}.

\subsection{DAOPHOT/ALLFRAME reduction: aperture and PSF fitting
photometry}
The  DAOPHOT/ALLFRAME reduction package  has  been extensively used by
the astronomical  community  and is   a  very well  tested   reduction
technique.  The idea behind  this stellar photometry package  consists
in modelling   the  PSF  of  each  image following  a  semi-analytical
approach, and in fitting  the derived model  to all  the stars in  the
image by means of least square method.   After some tests, we chose to
calculate a variable PSF across the field (quadratic variability).
We  selected the  first   $200$ brightest,  unsaturated stars in  each
frame, and calculated  a  first approximate PSF   from them.  We  then
rejected the stars to which  DAOPHOT assigned anomalously high fitting
$\chi$     values. After having  cleaned      the  PSF star list,   we
re-calculated the  PSF. This  procedure  was iterated  three times  in
order to obtain the final PSF for each image.

We  then used   DAOMATCH  and DAOMASTER   (Stetson  1992) in  order to
calculate the coordinate   transformations among the  frames, and with
MONTAGE2 we created a reference image by adding up  the 50 best seeing
images. 
We used this high S/N image to create a  master list of stars,
and  applied ALLFRAME  (Stetson  1994)  to refine  the estimated  star
positions  and magnitudes   in all  the frames.   
We applied  a first selection on the  photometric quality of our stars
by rejecting  all stars with SHARP and   CHI parameters (Stetson 1987)
deviating by more than 1.5 times the RMS of  the distribution of these
parameters from  their mean values,   both calculated in  bins of  0.1
magnitudes.  About  25\% of     the stars  were  eliminated by    this
selection.  This was  the PSF fitting   photometry we used in  further
analysis.    The  aperture photometry  with  neighbor subtraction was
obtained  with  a   new version of  the   PHOT   routine (developed by
P.~B.~Stetson).  We   used  as  centroids  the  same   values used for
ALLFRAME.   The  adopted apertures   were  equal to  the  FWHM  of the
specific image, and after some tests we set the annular region for the
calculation  of the sky   level  at a distance of  $1^{''}<r<2.5^{''}$
(both for the ALLFRAME and for the aperture photometry).

Finally, we  used again DAOMASTER   for the  cross-correlation of  the
final star lists, and to prepare the light curves.

\subsection{Image Subtraction photometry}
\label{s:imagesubtraction}

In the  last years  the  Image Subtraction  technique has been largely
used in photometric reductions. This method firstly implemented in the
software ISIS     did not    assume any specific functional  shape for
the PSF of  each image. Instead  it modeled the  kernel that convolved
the PSF of  the reference image to  match  the PSF of a  target image.
The reference  image is  convolved  by the   computed kernel and  then
subtracted from  the image.  The  photometry   is  then done  on   the
resulting difference image.
Isolated stars were  not required in  order to model the kernel.  This
technique had rapidly  gained an appreciable  consideration across the
astronomical  community. Since  its  advent, it appeared  particularly
well suited for the search for variable stars, and it has proved to be
very  effective in  extremely  crowded  fields like  in  the  case  of
globular clusters (e.g.   Olech  et  al.~\cite{olech99},   Kaluzny  et
al.~\cite{kaluzny01}, Clementini et al.~\cite{clementini04}, Corwin et
al.~\cite{corwin06}).
An extensive  use  of this  approach  has  been applied  also in  long
photometric  surveys devoted  to    the search for  extrasolar  planet
transits (eg. Mochejska et al., \cite{mochejska02,mochejska04}).

We  used the  standard reduction routines   in the ISIS2.2 package. At
first, the  images were  interpolated on the  reference system  of the
best  seeing image. Then we  created a reference  image  from the $50$
images with best seeing.
We performed different  tests in order to set  the best parameters for
the  subtraction, and we checked  the images to find which combination
 with lowest residuals.
In the end,  we decided to sub-divide the  images in four sub-regions,
and to apply a kernel and sky background variable at the second order.

  Using ALLSTAR, we build up  a master list of stars from the reference
image.
As  in Bruntt et al.~(\cite{bruntt03}),  we were not  able to obtain a
reliable photometry using the standard photometry routine of ISIS.  We
used the DAOPHOT aperture photometry routine  and slightly modified it
in order  to accept the subtracted  images. Aperture, and  sky annular
region  were  set as  for the  aperture and  PSF photometry.  Then the
magnitude of the stars in the subtracted images  was obtained by means
of the following formula:

\begin{equation}
\label{imsub}
m_i\,=\,m_{\rm ref}\,-\,2.5\,log\,\Big(\frac{N_{\rm ref} - N_i}{N_{\rm Ref}}\Big)
\end{equation}

where $m_i$ was the magnitude of a generic star  in a generic i$^{th}$
subtracted image, $m_{\rm ref}$ was the magnitude of the correspondent
star in the reference image, $N_{\rm ref}$ were the counts obtained in
the reference image and $N_i$ is the i$^{th}$ subtracted image.

\subsection{Zero point correction}
For  what concerns psf fitting photometry  and aperture photometry, we
corrected the  light  curves  taking  into   account the zero   points
      magnitude (the mean  difference in stellar magnitude) between  a
generic target image, and  the best seeing   image.  This was  done by
means of DAOMASTER and can be considered  as a first, crude correction
of the light curves.  Image subtraction was able to handle these first
order corrections automatically,  and thus the resulting light  curves
were already free of large zero points offsets.

Nevertheless,  important  residual  correlations     persisted  in the
light curves, and it was necessary to apply specific, and more refined
post-reduction corrections, as explained in the next Section.

\subsection{The post reduction procedure}
\label{s:postred}
In general,     and for   long time series photometric studies, it has
been   commonly recognized   that       regardless  of   the   adopted
reduction technique,    important correlations  between   the  derived
magnitudes and various parameters like seeing, airmass, exposure time,
etc.   persist in the final photometric sequences.
As put  into evidence by Pont et  al.~(\cite{pont06}), the presence of
correlated noise in real  light curves, (red noise), can significantly
reduce the   photometric precision  that  can  be obtained,  and hence
reduce transit detectability.
For example ground-based    photometric measurements are  affected  by
color-dependent atmospheric extinction. This is a problem
 since, in general, photometric surveys employ only one filter and
no explicit colour information is available.
To take  into account these effects,  we used the  method developed by
Tamuz et al.~(\cite{tamuz05}).  The software was  provided by the same
authors, and   an accurate  description  of it  can  be found  in  the
referred paper.

One of the  critical points in this algorithm  regarded the choice  of
systematic  effects to be  removed from the  data. For each systematic
effect,  the algorithm performed an  iteration  after which it removed
the systematic passing to the next.

To verify the     efficiency   of  this  procedure  (and establish the
number  of   systematic  effects to  be    removed) we  performed  the
 following simulations. 
We started from  a set of $4000$  artificial stars       with constant
magnitudes.   These      artificial  light  curves were created with a
realistic  modeling of  the noise  (which accounts  for  the intrinsic
noise of the  sources,  of the sky   background, and of  the  detector
electronics) but also for  the photometric reduction algorithm itself,
as described in Sec.~\ref{s:far}.
Thus, they are fully representative of the systematics present in our
data-set.
 At this point we also add $10$  light curves   including
transits      that were   randomly  distributed inside  the  observing
window.      These  spanned    a  range  of   (i)  depths from a   few
milli-magnitudes  to around $5$   percents,  (ii) durations  and  (iii)
periods  respectively  of  a few  hours  to  days  accordingly  to our
assumptionts on the  distributions  of these parameters  for planetary
transits, as accounted in Sec.~\ref{s:addtransit}.

     In a second step,   we applied the Tamuz et al.~(\cite{tamuz05})
algorithm     to the     entire   light   curves  data-set.  
For a deeper understanding of the total noise effects  and 
the transit  detection  efficiency, we  progressively increased   the
number of systematic  effects  to  be  removed  (eigenvectors in   the
analysis described by Tamuz 2005).
 Typically, we started with $5$ eigenvectors and increased it to $30$.

Repeated experiments showed no significant $RMS $ improvement after the
 ten iterations.
The final $RMS$ was $15\%$-$20\%$ lower than the original $RMS$. Thus,
 the number of eigenvectors was set to 10.  In  no case the added
transits were removed  from the light   curves and the transit  depths
remained unaltered.

We conclude that   this   procedure, while reducing   the  $RMS$
 and providing a very effective correction for  systematic effects,
 did not
influence the   uncorrelated  magnitude variations    associated  with
transiting planets.

\section{Definition of the photometric precision}
\label{s:photometricprecision}
To compare the performances of the different photometric algorithms we
calculated the    Root  Mean Square,   ($RMS$),  of   the  photometric
measurements obtained for each star, which is defined as:

\begin{equation}
RMS=\sqrt{\frac{\sum_{i=1}^{i=N}{\Big(\frac{I_{i}-<I>}{<I>}\Big)^{2}}}{N-1}}
\end{equation}

Where $I_{i}$ is the brightness measured in the generic $i^{th}$ image
for a particular target star, $<I>$ is the mean star brightness in the
 entire data-set, and $N$ is the number of images in which the star has been
measured. 
For constant stars, the relative variation of brightness is
 mainly due to the photometric measurement noise.  Thus,  the $RMS$
(as defined above) is equal to the mean  $N/S$ of the source.  
 In order to allow the detection of transiting jovian-planets
eclipses, whose relative brightness variations are of the order of
$1$\%, the $RMS$ of the source must be lower than this level.

\subsection{PSF photometry against aperture photometry}

\begin{figure*}
   \center
   \includegraphics[width=15cm]{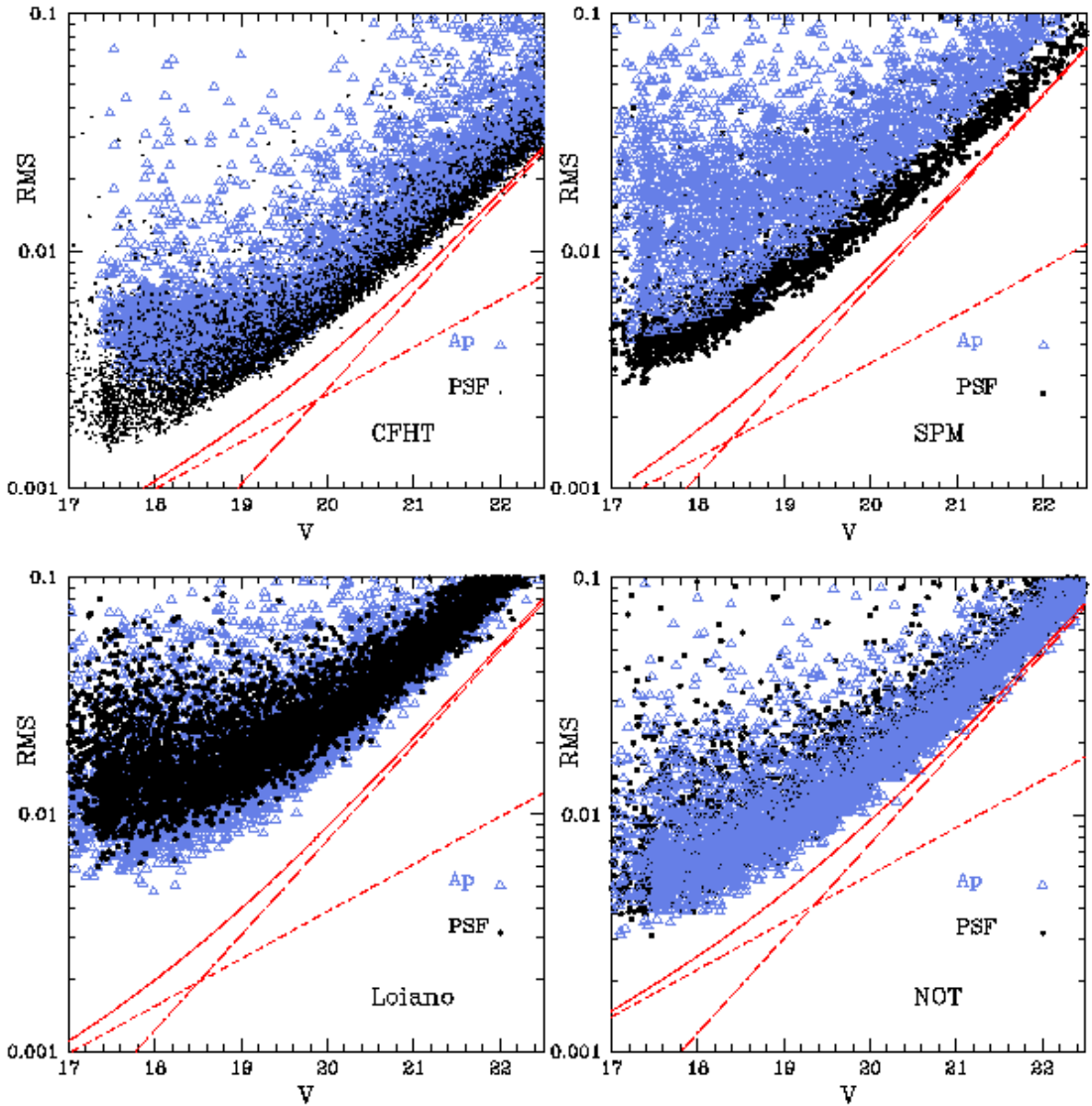}
      \caption{Comparison of the photometric precision for aperture photometry
               with neighbor subtraction (Ap) and for PSF fitting photometry
               (PSF) as a function of the apparent visual magnitude for
               CFHT, SPM, Loiano and NOT images.
               The short dashed line indicates the N/S considering
               only the star photon noise,
               the long dashed line is the N/S due to the sky photon
	       noise and the
	       detector noise. The continuous
               line is the total N/S.
              }
         \label{fig:confvspsf}
   \end{figure*}

The  first comparison  we  performed was       that between  aperture
photometry  (having   neighboring  stars   subtracted)  and PSF
fitting photometry.  We started with aperture photometry.
Figure~\ref{fig:confvspsf}, shows a comparison of the $RMS$ dispersion
of the light curves  obtained with the  new PHOT software with respect
to the $RMS$ of the PSF fitting photometry for the different sites.
We also show the theoretical noise which was estimated considering the
contribute to the noise coming from the photon noise of the source and
of  the sky    background as  well  as  the   noise  coming from   the
instrumental  noise   (see   Kjeldsen \&  Frandsen  \cite{kjeldsen92},
formula $31$).
In  Fig.~\ref{fig:confvspsf}-~\ref{rms},     we   also  separated  the
contribution of the source's Poisson noise from that of the sky (short
dashed line) and of the detector noise  (long dashed line).  The total
theoretical noise is  represented as a  solid line.  It  is clear that
the data-sets from the different telescopes gave different results.
In the case  of the CFHT and  SPM data-sets,  aperture photometry does
not reach the same  level of precision  as PSF fitting photometry (for
both bright and the faint sources).
Moreover, it appears  that the $RMS$  of aperture photometry reaches a
constant value below $V\sim18.5$ for CFHT  data and around $V\sim17.5$
for SPM data, while for PSF fitting  photometry the $RMS$ continues to
decrease.
 For Loiano data,  and  with  respect  to  PSF photometry, aperture
photometry provides a smaller $RMS$ in the  light  curve, in particular
for bright sources.    The  NOT  observations  on   the other hand show
that the two techniques are almost equivalent.
Leaving     aside    these   differences,  it   is   clear   that  the 
CFHT provide the  best photometric precision, and  this is due  to the
larger telescope  diameter, the  smaller telescope   pixel scale
($0.206$ arcsec/pixel,  see  Table~\ref{tab:6791}), and   the   better
detector performances at the CFHT.
For this data-set, the  photometric error remains smaller than  $0.01$
mag,   from  the turn-off of   the  cluster  ($\sim\,17.5$) to  around
magnitude $V=21$,  allowing the search  for transiting planets  over a
magnitude range of about $3.5$ magnitudes, (in fact, it is possible to
go one    magnitude deeper because of   the  expected increase  in the
transit     depth    towards    the fainter     cluster    stars,  see
Sec.~\ref{s:selection} for more details).
Loiano  photometry barely   reaches 0.01$\,$mag  photometric precision
even for the  brightest stars, a  photometric quality too poor for the
purposes  of our investigation.  The search  for planetary transits is
limited to  almost $2$ magnitudes below  the turn-off for SPM data (in
particular with  the  PSF fitting technique)  and to   $1.5$ magnitude
below  the  turn-off for the  NOT data.  In any  case, the photometric
precision for  the SPM and NOT  data-sets reaches the $0.004$ mag level
for the brightest  stars, while, for the  CFHT, it reaches the $0.002$
mag level.

\begin{figure}
   \center
   \includegraphics[width=8cm]{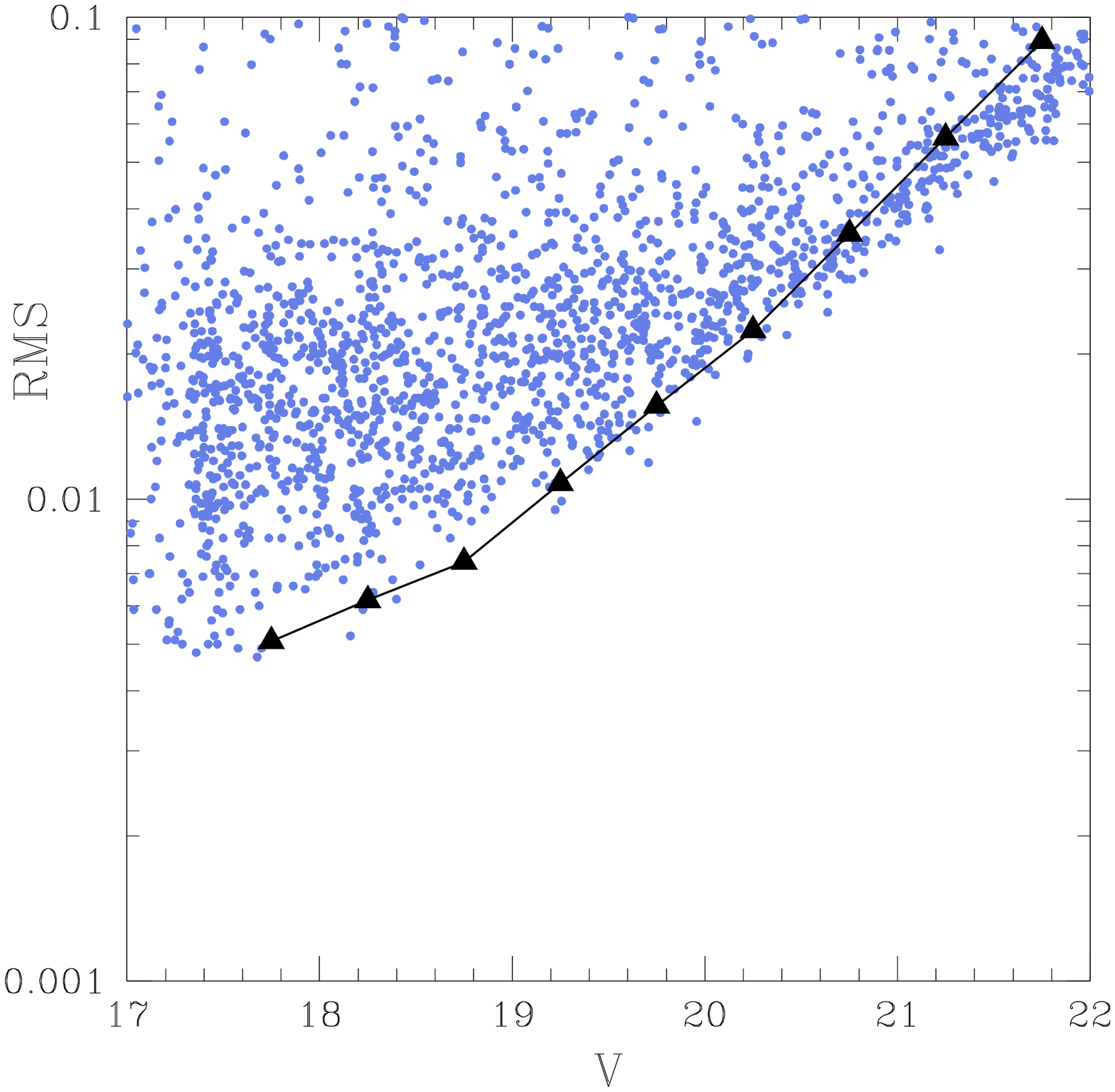}
      \caption{RMS noise for the San Pedro M\'artir observations with the
               aperture error (triangles) as estimated by Equation
               \ref{aperror}.
              }
      \label{varsky}
   \end{figure}

It is clear  that both  the PSF  fitting  photometry and the  aperture
photometry   tend to have  larger errors        with  respect to   the
expected error level.  This effect is much clearer for Loiano, SPM and
NOT rather than for Hawaii photometry.
As explained by  Kjeldsen  \& Frandsen (\cite{kjeldsen92}), and   more
recently  by   Hartman et   al.~(\cite{hartman05}), the  PSF   fitting
approach in general  results  in poorer photometry  for  the brightest
sources  with respect   to  the aperture   photometry.  But, for   our
data-sets, this was true  only for the  case  of Loiano  photometry, as
demonstrated above.
The  aperture  photometry routine in DAOPHOT   returns  for each star,
along  with other information,  the  modal sky value associated  with
that star, and  the ${\emph rms}$  dispersion ($\sigma_{\rm  sky}$) of
the sky  values   inside the sky   annular region.  So, we  chose   to
calculate the error associated with the random noise inside the star's
aperture with the formula:

\begin{equation}
\label{aperror}
\sigma_{\rm Aperture}=\sqrt{\sigma_{\rm sky}^2\,{\rm Area}}
\end{equation}

\noindent
where Area is the area (in pixels$^{2}$) of the region inside which we
measure the star's  brightness.   This error automatically  takes into
account the sky Poissonian noise,  instrumental effects like the  Read
Out  Noise,   (RON),    or detector  non-linearities,    and  possible
contributions of neighbor stars.
To calculate this error, we  chose a representative mean-seeing image,
and    subdivide      the   magnitude    range     in  the    interval
$17.5\,<\,V\,<\,22.0$ into nine bins of 0.5 mag.
Inside each of these bins, we took the minimum value of the stars' sky
variance  as representative of the sky  variance of the  best stars in
that magnitude bin.
We over-plot this  contribution in Fig.~\ref{varsky}  which is relative
to  the San Pedro photometry.  This  error completely accounts for the
observed  photometric precision.  So, the   variance inside the star's
aperture is much larger than what is expected from simple photon noise
calculations.  This  can be    the  effect of  neighbor stars   or of
instrumental problems.
For CFHT photometry, as we have good seeing  conditions and an optimal
telescope scale, crowding plays a less important role.  Concerning the
other sites, we noted that for Loiano the  crowding is larger than for
SPM and NOT, and this could explain the lower photometric precision of
Loiano observations, along with the smaller telescope diameter.
For  NOT photometry, instead,  the crowding should  be larger than for
San Pedro  (since the  scale of the   telescope  is larger)  while the
median   seeing conditions are comparable for    the two data-sets, as
shown in Sect~\ref{s:observations}.
Therefore this effect should be more evident for NOT      rather  than
for SPM, but this is not the case, at least for what concerns aperture
photometry.        For   PSF  photometry, as     it       is  seen  in
Fig.~\ref{fig:confisispsf}, the NOT photometric precision appears more
scattered than the SPM photometry.

We are  forced  to     conclude  that poor  flat    fielding,  optical
distortions,   non-linearity   of  the   detectors  and/or presence of
saturated pixels in the brightest stars must have played a significant
role
in setting the      aperture and PSF fitting    photometric precisions.

\subsection{PSF photometry against image subtraction photometry}

\begin{figure*}
   \center
   \includegraphics[width=15cm]{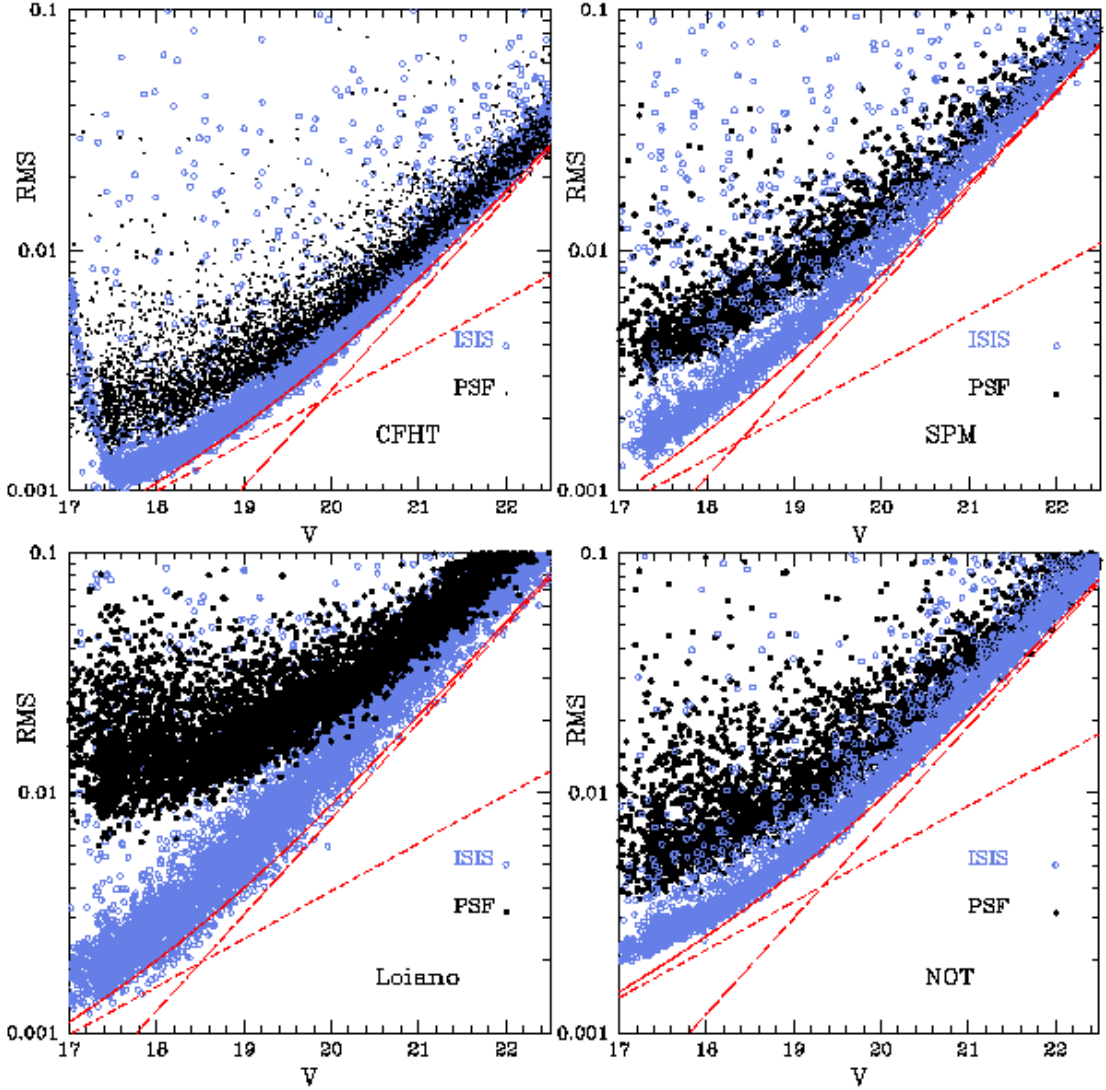}
      \caption{Comparison of the photometric precision for image
               subtraction (ISIS) and for psf fitting photometry
               (PSF) as function of the apparent visual magnitude for
	       CFHT, SPM, Loiano, and NOT images.
              }
         \label{fig:confisispsf}
   \end{figure*}

Applying  the image subtraction technique  we were able to improve the
photometric precision  with respect to that obtained   by means of the
aperture photometry and the PSF fitting techniques.
This appears evident in Fig.~\ref{fig:confisispsf}, in which the image
subtraction technique is compared to the PSF fitting technique.
Again, the best overall photometry  was  obtained for the $CFHT$,  for
the  reasons explained   in the previous   subsection. For  the  image
subtraction reduction, the  photometric precision overcame the $0.001$
mag level for the brightest  stars in the  $CFHT$ data-set, and for the
other sites  it was around $0.002$ mag  (for the $NOT$) or better (for
$SPM$ and  $Loiano$).  This clearly allowed  the search for planets in
all these different data-sets.
In this case, it was possible to  include also the Loiano observations
(up to $2$  magnitudes below the turn-off),  and, for the other sites,
to extend  by about $0.5$-$1$ mag, the  range of magnitudes over which
the search for transits was possible, (see previous Section).

The reason for which  image subtraction gave  better results could  be
that it  is more suitable for  crowded regions (as  the center of the
cluster), because it doesn't need  isolated stars in order to calculte
the convolution kernel while the subtraction  of stars by means of PSF
fitting can  give rise  to higher residuals,    because it's much  more
difficult to obtain a reliable PSF from crowded stars.

\subsection{The best photometric precision}

   Given  the results of the previous comparisons, we decided to adopt
the   photometric   data set  obtained    with  the  image subtraction
technique. Figure~\ref{rms}, shows  the photometric precision that  we
obtained for the four different  sites.  The photometric precision  is
very close to the theoretical noise for all the data-sets.
The NOT data-set has a lower photometric  precision with respect to SPM
and even to Loiano, in particular for the brightest stars. We observed
that the mean $S/N$ for  the NOT images  is  lower than for the  other
sites because of the  larger number of  images taken (and consequently
of their lower exposure times and $S/N$), see Tab.~\ref{tab:6791}.

   \begin{figure*}
   \center
         \label{rms}
   \includegraphics[width=15cm]{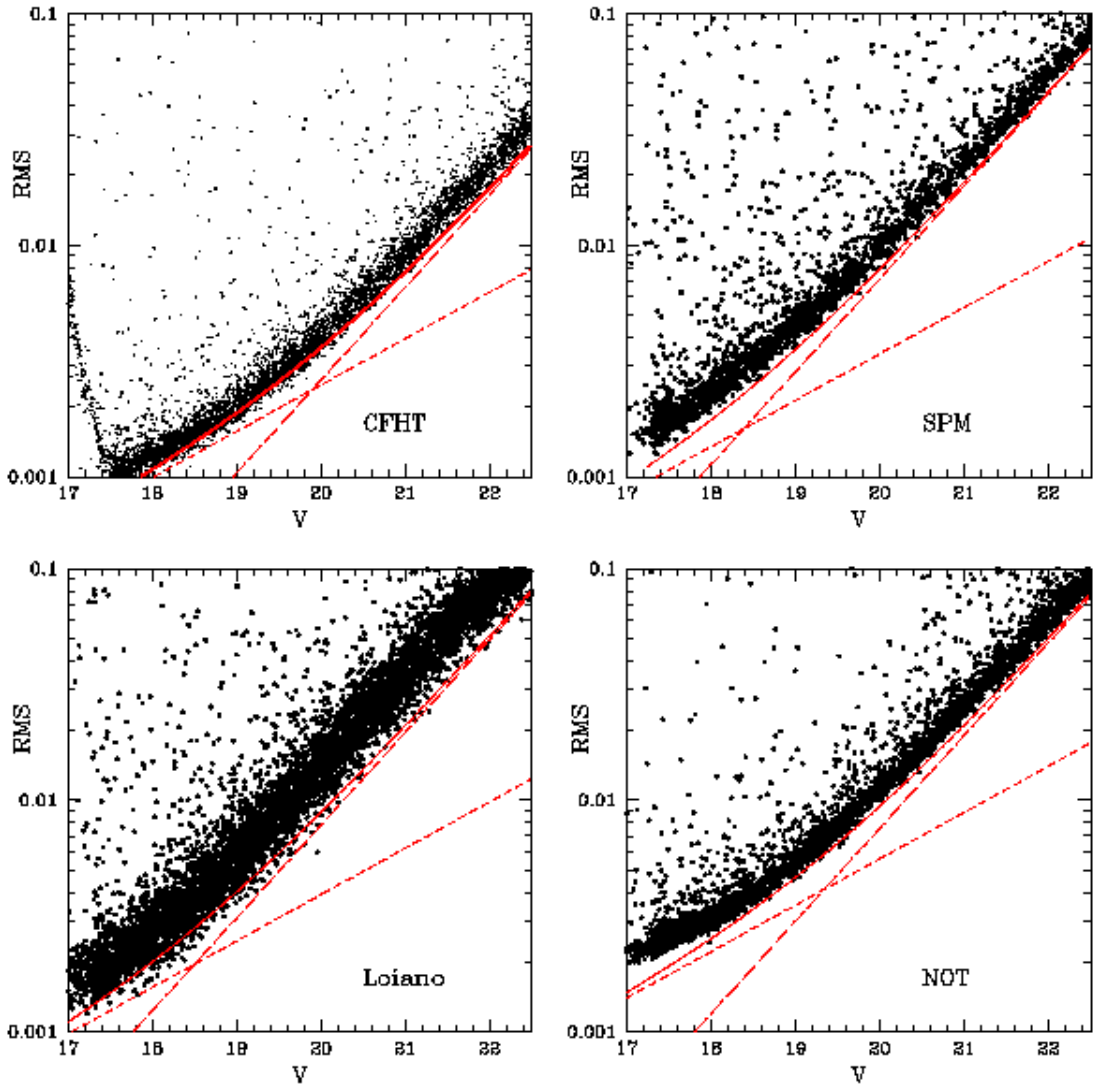}
      \caption{The expected RMS noise for the observations taken at
                the different sites as a function
                of the visual apparent magnitude, is compared with the
                RMS of the observed light curves obtained with the image
		subtraction technique.
               }
         \label{rms}
   \end{figure*}

\section{Selection of cluster members}
\label{s:selection}

\begin{figure}
   \center
   \includegraphics[width=8cm]{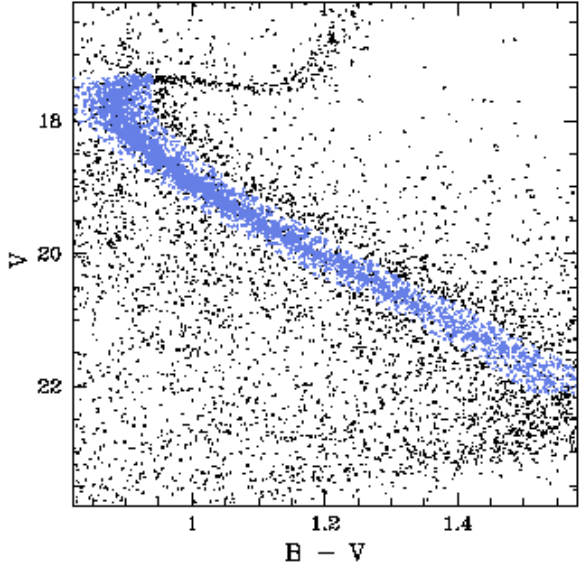}
      \caption{The  \object{NGC~6791} CMD   highlighting the selection
               region of the main sequence stars (blue circles).
              }
         \label{fig:selezioneMS}
   \end{figure}

 To detect   planetary transits in \object{NGC~6791}
we selected the probable main sequence cluster members as follows.
Calibrated magnitudes  and colors  were obtained by  cross-correlating
our   photometry    with    the      photometry  by  Stetson        et
al.~(\cite{stetson03}), based  on the same  NOT  data-set  used in the
present paper.  Then, as     done   in M05, we considered $24$ bins of
$0.2$ magnitudes in the interval $17.3\,<\,V\,<\,22.1$.
For each bin,   we calculated a robust  mean  of  all  {\em(B-V)} star
colors,   discarding  outliers with   colors differing   by  more than
$\sim0.06$ mag from the mean.   Our selected main sequence members are
shown in Fig.~\ref{fig:selezioneMS}.    Overall,  we selected   $3311$
main-sequence candidates in
\object{NGC~6791}.
These are the stars present in at least  one of the four data-sets (see
Sec.~\ref{s:diffapproach}),      and  represent the candidates for our
planetary transits search.

Note that our selection criteria excludes stars in the binary sequence
of the cluster.  These  are blended objects,   for  which any  transit
signature  should  be   diluted by    the   light of  the   unresolved
companion(s) and then likely undetectable.
Furthermore, a narrow selection range helps in reducing the field-star
contamination.

\section{Description of the transit detection technique}
\label{s:algorithm}

\subsection{The box fitting technique}

To       detect   transits in our   light  curves we  adopted the  BLS
algorithm by   Kov\'acs et  al.~(\cite{kovacs02}).   This technique is
based on the fitting of a box shaped transit model to the data.
It assumes that the value of the magnitude  outside the transit region
is constant. It is  applied to the  phase folded  light curve of  each
star spanning  a range of  possible orbital periods for the transiting
object, (see Table~\ref{tab:parametri}).  Chi-squared minimization  is
used to obtain the best model solution.
The quantity to be maximized in order to get the best solution is:

\begin{equation}
T = \Big(\sum_{n=in}m_n\Big)^2\Big[\frac{1}{N_{in}\,N_{out}}\Big]
\end{equation}

\noindent
where  $m_n=M_n\,-\,<M>$.   $M_n$  is the $n$-th    measurement of the
stellar  magnitude in the light curve,  $<M>$ is the mean magnitude of
the  star and   thus   $m_n$ is   the  n-th residual   of the  stellar
magnitude.   The  sum at    the  numerator  includes all   photometric
measurements that fall inside the transit region.
Finally $N_{in}$  and  $N_{out}$    are respectively the   number   of
photometric measurements inside and outside the transit region.

The algorithm,
at first, folds the light curve assuming a particular period.
 Then, it sub-divides the folded light curve in $nb$ bins
and starting from each one of these bins calculates the $T$ index
shown above spanning a range of transit lengths between $qmi$
and $qma$ fraction of the assumed period. 
Then, it provides the period,  the depth of the brightness  variation,
$\delta$, the  transit length, and the initial  and final bins  in the
folded light curve at which the maximum value of the index $T$ occurs.
We used a routine called 'eebls'
available                            on                            the
web\footnote{http://www.konkoly.hu/staff/kovacs/index.html}.        We
applied   also  the   so    called   directional correction   (Tingley
\cite{tingley03a},  \cite{tingley03b}) which  consists in taking  into
account the  sign of the  numerator in the above  formula in  order to
retain only the brightness variations which imply a positive increment
in apparent magnitude.

\subsection{Algorithm parameters}

The parameters to be set     before running    the BLS algorithm are
the following:
1) {\em nf}, number of frequency points for which the spectrum is computed;
2) {\em fmin}, minimum frequency;
3) {\em df}, frequency step;
4) {\em nb}, number of bins in the folded time series at any test frequency;
5) {\em qmi}, minimum fractional transit length to be tested;
6) {\em qma}, maximum fractional transit length to be tested;
$qmi$ and $qma$ are given as the product of the  transit length to the
test   frequency.    Table~\ref{tab:parametri}  displays  our  adopted
parameters.

\begin{table}
\caption{Adopted parameters for the BLS algorithm: nf is the
number of frequency steps adopted, fmin is the minimum frequency
considered, df is the increasing frequency step, nb is the number
of bins in the folded time series at any test frequency, qmi and qma
are the minimum and maximum fractional transit length to be tested,
as explained in the text.
\label{tab:parametri}
}

\begin{center}
\begin{tabular}{ c c c c c c}
\hline
   nf & fmin(days$^{-1}$) & df(days$^{-1}$) & nb & qmi & qma  \\
\hline
 3000 & 0.1 & 0.0005 & 1000 & 0.01 & 0.1 \\
\hline
\end{tabular}
\end{center}
\end{table}

\subsection{Algorithm transit detection criteria}
\label{s:threshold}
To characterize the statistical significance  of a transit-like  event
detected by the BLS algorithm we  followed     the methods by Kov\'acs
\&  Bakos (\cite{kovacs05}):  deriving the Dip  Significance Parameter
(hereafter DSP) and the significance of the main  period signal in the
Out of Transit  Variation (hereafter  OOTV, given  by  the folded time
series with the exclusion of the transit).

The Dip Significance Parameter is defined as

\begin{equation}
{\rm    DSP}        =      \delta(\sigma^{2}/N_{\rm     tr}+A^{2}_{\rm
OOTV})^{-\frac{1}{2}}
\end{equation}

\noindent
where  $\delta$ is the  depth of the  transit given by  the BLS at the
point  at which the index  $T$  is maximum,  $\sigma$  is the standard
deviation  of the $N_{\rm tr}$ in-transit  data points, $A_{\rm OOTV}$
is  the peak amplitude  in the Fourier spectrum  of the Out of Transit
Variation.
The threshold for the DSP set by  Kovacs \& Bakos (\cite{kovacs05}) is
$6.0$ and it was set on artificial constant light curves with gaussian
noise. In real light curves the noise is not gaussian, as explained in
Sec.~\ref{s:postred}, and, in general, the value  of the DSP threshold
should be set case by case. In Sec.~\ref{s:diffapproach}, we presented
the adopted  thresholds, based on  our simulations on artificial light
curves, described in Sec.~\ref{s:simul}.

The significance of the main periodic signal in the OOTV is defined as:

\begin{equation}
{\rm SNR}_{\rm OOTV}=\sigma_{A}^{-1}(A_{\rm OOTV}-<A>)
\end{equation}

\noindent
where $<A>$  and   $\sigma_{A}$  are  the average and     the standard
deviation  of the Fourier spectrum.    This parameter accounts for the
Out Of Transit Variation, and we impose it to  be lower than $7.0$, as
in Kovacs \& Bakos (\cite{kovacs05}).

For our search we imposed a maximum transit  duration of six hours; we
also required that  at least ten data  points must be  included in the
transit region.
%


\section{Simulations}
\label{s:simul}
The Transit  Detection Efficiency (TDE) of  the adopted  algorithm and
its False   Alarm Rate  (FAR) were  determined   by means  of detailed
simulations.  The  TDE is   a  measure of   the probability  that   an
algorithm correctly identifies a transit in a light  curve. The FAR is
a measure of the probability that an algorithm identifies a feature in
a light curve that does not represent a transit, but rather a spurious
photometric effect.

In  the following  discussion,      we  address    the details  of the
simulations  we performed,       considering   the case    of the CFHT
observations of \object{NGC~6791}.  Because the CFHT data provided the
best  of  our  photometric sequences,  the  results  on  the algorithm
performance is shown below, and should be considered as an upper limit
for the other cases.

\subsection{Simulations with constant light curves}
\label{s:far}

\begin{figure}[!]
   \center
   \includegraphics[width=8cm]{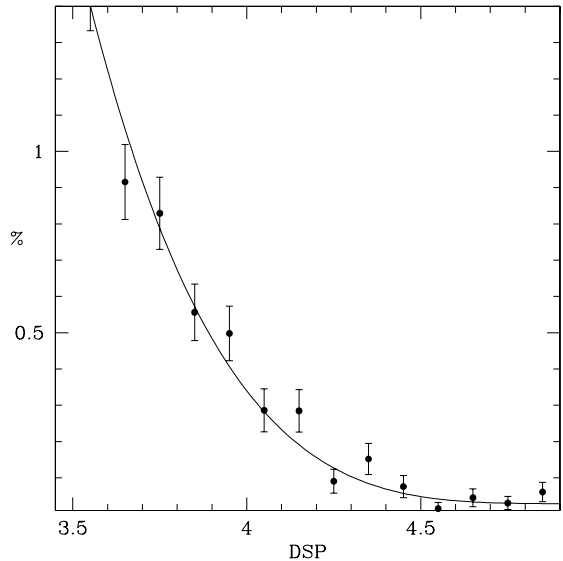}
      \caption{ False Alarm Probability (FAR) in \%, against the DSP
	       parameter
               given by the algorithm. The points indicate the results
               of our simulations on constant light curves, the
               solid line is our assumed best fit.  }
         \label{fig:effcost}
   \end{figure}

Artificial stars with   constant magnitude were  added to  each image,
according to    an equally-spaced grid   of  2*PSFRADIUS+1, (where the
PSFRADIUS was the region  over which the  Point Spread Function of the
stars was calculated, and was around $15$ pixels for the CFHT images),
as described in Piotto \& Zoccali (\cite{zoccali}).
We took into account  the photometric zero-point differences among the
images, and the coordinate transformations  from one image to another.
$7722$  stars were added on  the CFHT images.  In  order to assure the
homogeneity  of these simulations,  the   artificial stars were  added
exactly  in the same  positions,  (relative to  the real stars  in the
field), for the other sites.  Because  of the different field of views
of  the detectors, (see  Tab.~\ref{tab:6791}), the number of resulting
added stars was $3660$ for the NOT, $5544$  for Loiano, and $3938$ for
SPM.
The     entire   set  of images   was   then reduced again  with   the
procedure described  in Sec.\ref{s:reduction}.  This way  we got a set
of constant light curves which is completely representative of many of
the   spurious   artifacts that could   have   been introduced  by the
photometry procedure.
This is certainly a    more  realistic test than simply    considering
Poisson noise on  the light  curves, as it  is  usually done.  We  then
applied  the  algorithm,      with  the   parameters described      in
Sec.\ref{s:algorithm}, to the constant light curves.
The  result is  shown   in  Figure~\ref{fig:effcost}, where  the   DSP
parameter  is plotted   against    the mean magnitude   of  the  light
curve. For the  CFHT data, fixing the  DSP threshold at 4.3  yielded a
FAR of $0.1\%$. This was the FAR  we adopted also when considering the
other  sites,  which corresponded   to   different levels of the   DSP
parameter, as explained in Sec.~\ref{s:diffapproach}.

Repeating the   whole procedure $4$ times  and  slightly  shifting the
positions of the  artificial stars, allowed  us to better estimate the
FAR and  its  error,  FAR=$(0.10\pm0.04)\%$.   Therefore, running  the
transit search procedure on the  $3311$ selected main sequence  stars,
we expect $(3.3\,\pm\,1.3)$ false candidates.

\subsection{Masking bad regions and temporal intervals}
\label{s:screening}
We verified that, when stars were located  near detector defects, like
bad  columns, saturate stars,  etc.,  or,  in correspondence of   some
instants  of a  particular   night, (associated  with  sudden climatic
variations, or  telescope  shifts),   it  was  possible to   have   an
over-production of   spurious transit   candidates.   To  avoid  these
effects,  we chose to mask   those regions of   the detectors and  the
epochs which caused sudden changes in the photometric quality.
This was done also for the simulations with  the constant added stars,
that were not inserted in detector defected regions, and in the
 excluded images that   generating  bad  photometry. In particular, we
observed these spurious effects for the NOT and SPM images.
We further observed,  when  discussing the candidates coming  from the
analysis of the whole data-set (as described in Sec.\ref{s:whole}) that
    the photometric variations    were  concentrated   on    the first
 night of the NOT.
This fact, which appeared from the simulations with the constant stars
too,  meant that this night  was probably      subject to  bad weather
conditions.
had  not  we applied  the
Because  we  didn't recognize it  at  the beginning, we  retained that
night,  as long  as  those candidates,   which were all  recognized of
spurious nature.  Had  not  we applied  any
masking the number of false alarms would have almost quadruplicated.
This fact  probably can explain at least  some of the candidates found
by  B03 (see Sec.~\ref{s:comparison}) that were  identified on the NOT
observations.
Even  if  some kind of masking   procedure  was applied  by  B03, many
candidates  appeared   concentrated  on the     same  dates, and  were
considered rather suspicious by the same authors.

\subsection{Artificially added transits}
\label{s:addtransit}
The transit  detection efficiency (TDE)   was determined by  analyzing
light  curves modified with  the inclusion  of   random transits.   To
properly measure the  TDE and to  estimate  the number of  transits we
expect to   detect it is  mandatory   to consider realistic planetary
transits.  We proceeded as follows:

\subsubsection{Stellar parameters}

\begin{figure}
   \center
    \includegraphics[width=0.9\columnwidth]{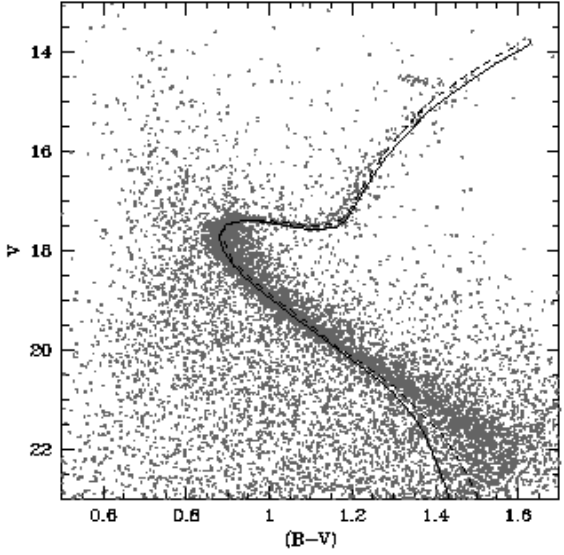}
      \caption{\footnotesize CMD diagram of~\object{NGC6791} with the 
	best fit Z=0.030
	isochrone (dashed line), and the best fit Z=0.046 isochrone
	(from Carraro et al.~\cite{carraro06}, solid line).
	Photometry: Stetson et al.~(\cite{stetson03}) }
        \label{hr}
   \end{figure}

The  basic cluster parameters   were determined by fitting theoretical
isochrones  from  Girardi et   al.~(\cite{girardi02}) to the  observed
color-magnitude  diagram (Stetson  et al.~\cite{stetson03}).  Our best
fit parameters are (see Fig.~\ref{hr}): age = 10.0 Gyr, $(m-M)=13.30$,
$E_{\rm (B-V)}=0.12$  for $Z=0.030$ (corresponding to [Fe/H]$=+0.18$),
and age = 8.9 Gyr, $(m-M)=13.35$  and $E_{\rm (B-V)}=0.09$ for Z=0.046
(corresponding to [Fe/H]$=+0.39$).

\begin{figure}
   \center
    \includegraphics[width=0.49\columnwidth]{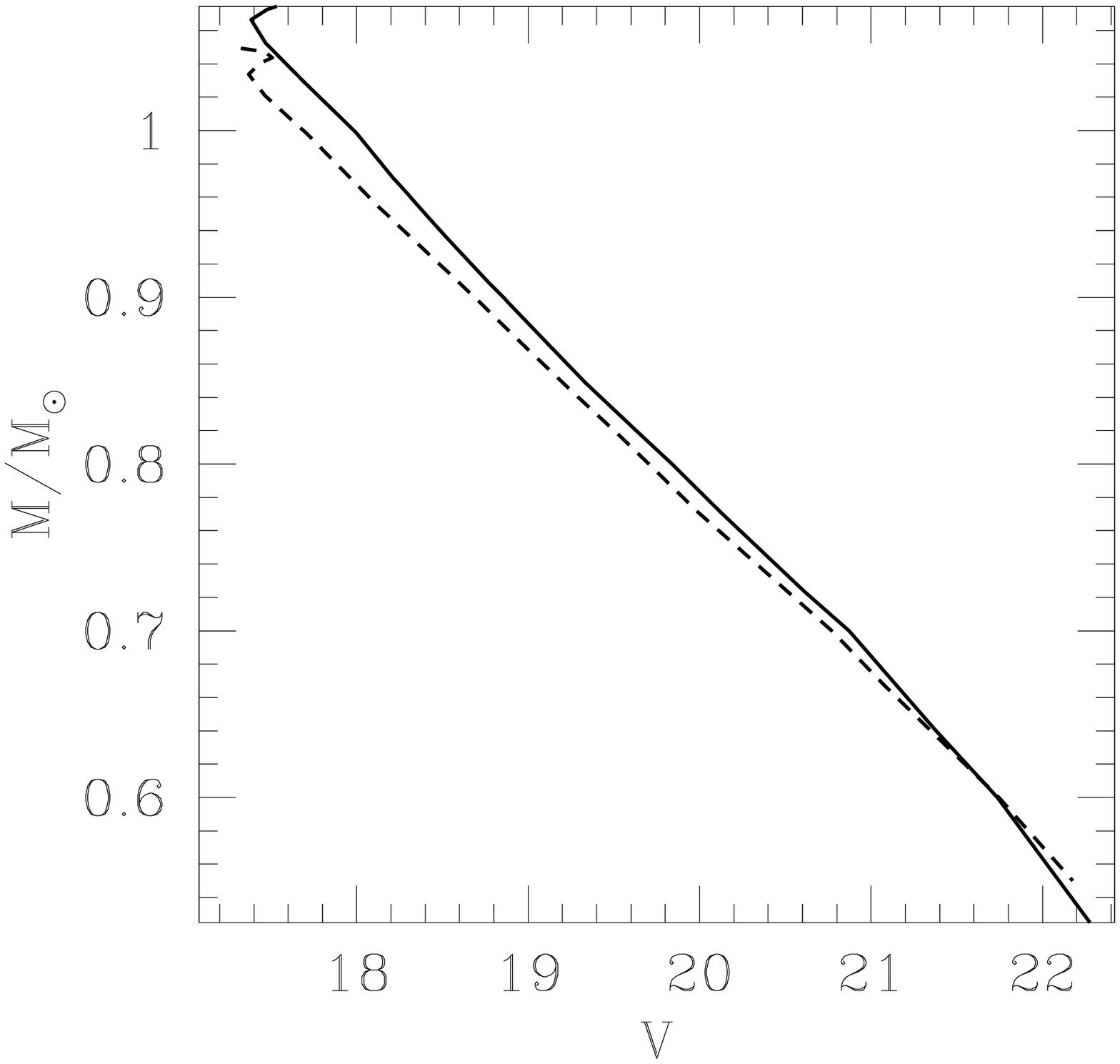}
    \includegraphics[width=0.49\columnwidth]{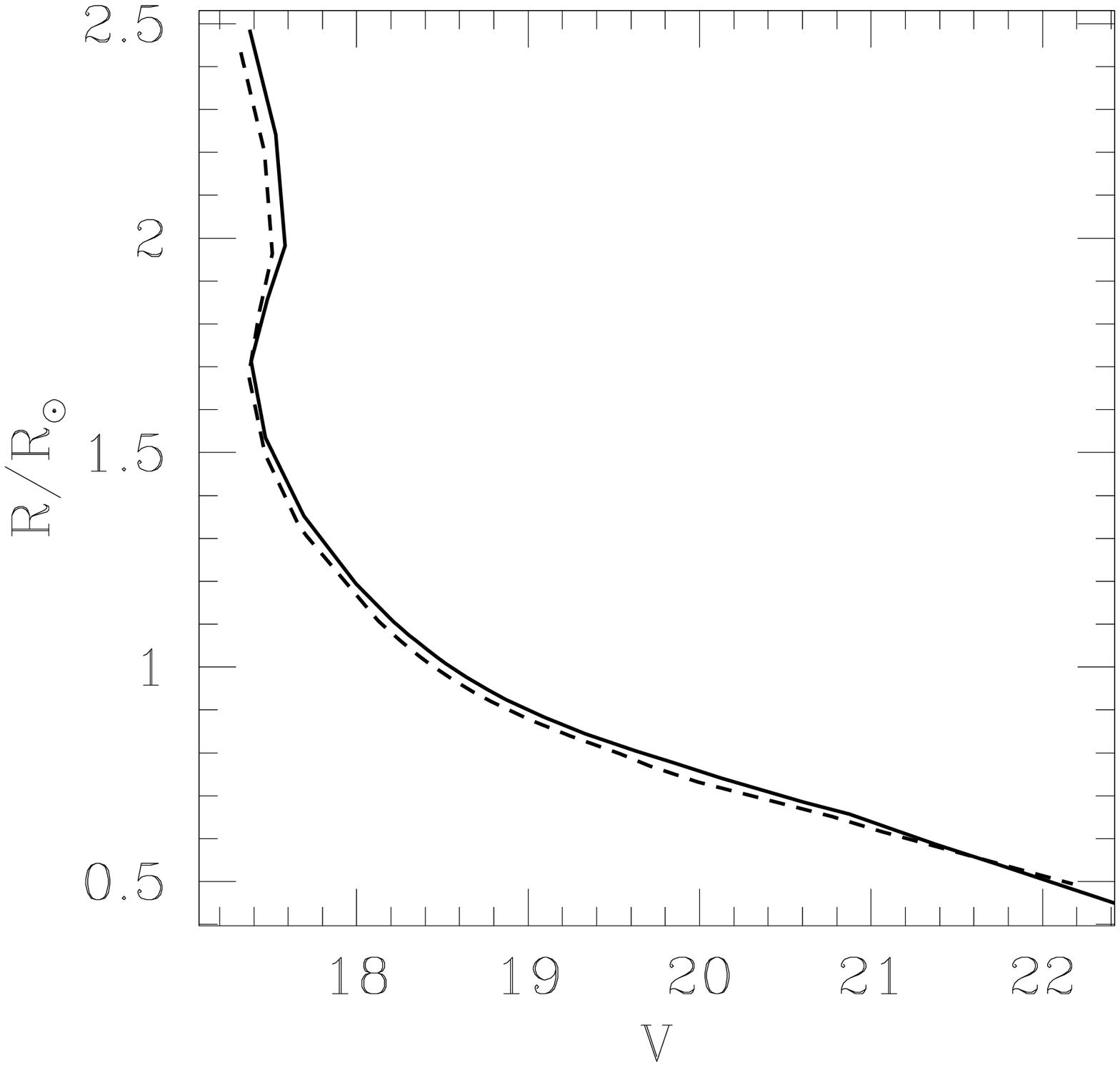}
      \caption{\footnotesize {\em Left}: $M_{i}/M_{\odot}$ vs visual 
	         apparent magnitude {\em Right}:
                 $R_{i}/R_{\odot}$ vs visual apparent magnitude, from
                 our best fit isochrone (dashed line)
                 and from the Z=0.046 isochrone (solid line)
		 applied to the stars of~\object{NGC~6791}.}
        \label{fig_vmr}
   \end{figure}

From the best-fit  isochrones we then  obtained the values of  stellar
mass   and radius    as    a  function  of  the     visual   magnitude
(Fig.~\ref{fig_vmr}).

\subsubsection{Planetary parameters}
\label{s:planetpar}
The actual distribution of planetary radii has a very strong impact on
the transit depth and therefore on the number of planetary transits we
expect to be able to detect.
The radius   of the  fourteen  transiting planets   discovered to date
ranges        from R=$1.35\,\pm\,0.07\,R_{J}$     (\object{HD209458b};
Wittenmyrer et al.~\cite{wittenmyer05}) to R=$0.725\,\pm\,0.03\,R_{J}$
(\object{HD149026b}; Sato et  al.~\cite{sato05}), where $J$ refers  to
the value for Jupiter.
The observed  distribution   is likely biased   towards  larger radii.
Gaudi (2005a)  suggests for the close-in giant   planets a mean radius
$R_{p}=1.03~\,~R_{J}$.  To evaluate the efficiency of the algorithm we
have considered three cases:
\begin{itemize}
\item $R_{p}=(0.7\,\pm\,0.1)\,R_{J}$
\item $R_{p}=(1.0\,\pm\,0.2)\,R_{J}$
\item $R_{p}=(1.4\,\pm\,0.1)\,R_{J}$
\end{itemize}

\noindent assuming a Gaussian distribution for $R_{p}$.
We fixed the planetary mass at $M_{p}=1\,M_{J}$, because the effect of
planet mass on transit depth or duration is negligible.

\begin{figure}
   \center
    \includegraphics[width=0.9\columnwidth]{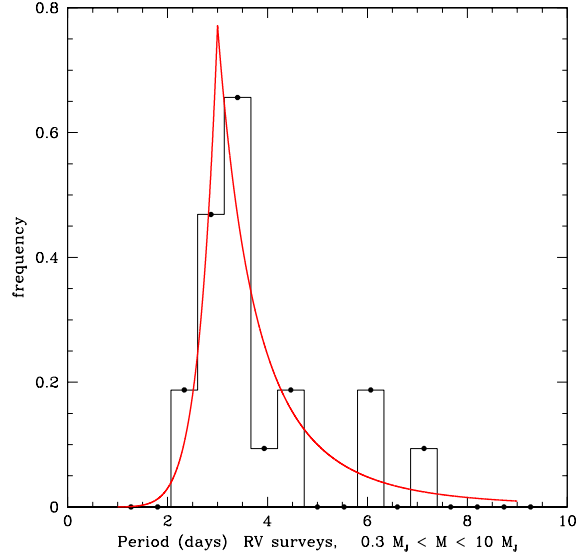}
      \caption{\footnotesize {\em Continuous line}: adopted distribution for
	planet periods. {\em Histogram}:
        RV surveys data (from the Extrasolar Planets Encyclopaedia).}
        \label{fig_period}
   \end{figure}

The period distribution was taken from the data for planets discovered
by  radial   velocity     surveys,  from  the     Extra-solar  Planets
Encyclopaedia\footnote{http://exoplanet.eu/index.php}.
We  selected  the planets discovered  by  radial velocity surveys with
mass $0.3 M_{J} \leq M_{pl} \sin i \leq 10 M_{J}$ (the upper limit was
fixed  to exclude brown dwarfs; the  lower limit  to ensure reasonable
completeness of RV surveys and to exclude Hot Neptunes that might have
radii      much  smaller   than     giant     planets,   Baraffe    et
al.~\cite{baraffe05}) and periods $1 \leq P \leq 9$ days.
We assumed that the  period distribution of RV  planets is unbiased in
this period range.  We  then  fitted the observed period  distribution
with a  positive power law for  the Very Hot Jupiters  (VHJ, $1 \leq P
\leq 3$)  and a negative power  law for the Hot Jupiters   (HJ, $3 < P
\leq    9$, see   Gaudi   et  al.~2005b  for    details)  as shown  in
Fig.~\ref{fig_period}.

\subsubsection{Limb-darkening}

To obtain realistic transit curves it is important to include the limb
darkening  effect.  We  adopted a   non-linear  law for  the  specific
intensity of a star:

\begin{equation}
\frac{I(\mu)}{I(1)}=1-\sum_{k=1}^{4}\,a_{k}\,(1-\mu^{k/2})
\end{equation}

\noindent
from Claret (\cite{claret}).

In this relation $\mu=\cos \gamma $ is the cosine of the angle between
the  normal to the   stellar surface  and  the  line  of sight  of the
observer, and $a_k$  are   numerical  coefficients that  depend   upon
$v_{turb}$  (micro-turbulent  velocity),  [M/H],  $T_{eff}$,  and  the
spectral band.
The coefficients are  available from the ATLAS calculations (available
at CDS).

We  adopted the metallicity of the  cluster for [M/H] and $v_{turb}$=2
km s$^{-1}$  for   all  the  stars.  For each   star  we  adopted  the
appropriate $V$-band $a_k$ coefficients as a function of the values of
$\log g$ and $T_{eff}$ derived from the best fit isochrone.

\subsubsection{Modified light curves}

In order to  establish  the TDE  of  the algorithm, we  considered the
whole sample of constant stars with $17.3 \leq  V \leq 22.1$, and each
star     was assigned   a planet with mass, radius and period randomly
selected  from   the  distributions   described  above.  The   orbital
semi-major axis   $a$  was derived from   the   $3^{rd}$ Kepler's law,
assuming circular orbits.

To each planet,  we also assigned  an orbit with a random  inclination
angle $i$,  with $0 < \cos i  < 0.1 $,  with a uniform distribution in
$\cos i$.   We      infer that    $\sim85\%$ of the planets
 result in potentially detectable transits.   We  also   assigned a phase
$2\phi_0$ randomly chosen  from  $0$  to  $2  \pi$  rad and a   random
direction of revolution $s = \pm 1$ (clockwise or counter-clockwise).

\begin{figure}
   \center
    \includegraphics[width=0.9\columnwidth]{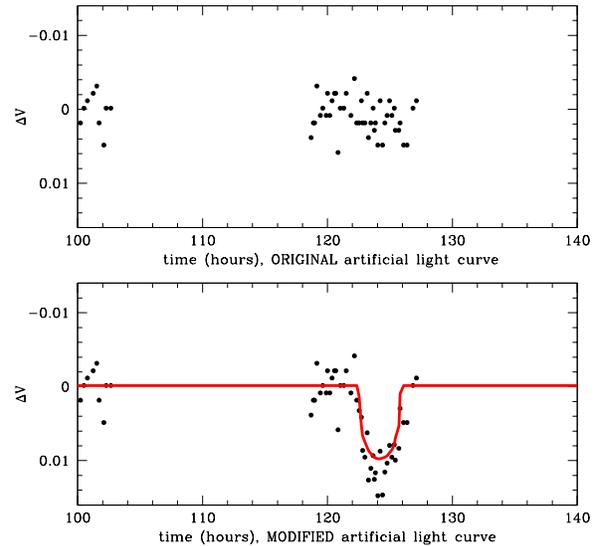}
      \caption{\footnotesize {\em Top:} constant light curve
       {\em Bottom}: the same light curve after
      inserting the simulated transit with limb-darkening (black points). The
      {\em solid line} shows the theoretical light curve of the transit.}
        \label{figtransit}
   \end{figure}

 Having fixed the planet's parameters ($P$, $i$, $\phi_0$, $M_p$, $R_p$, $a$),
the  star's  parameters ($M_\star$, $R_\star$)    and a constant light
curve ($t_i$\,, $V_i$) it is now possible to  derive the position
of  the planet with  respect  to  the  star  at  every instant
 from  the relation:
\[
\phi = \phi_0 + \frac{2 \pi s}{P}t_i
\]

\noindent
where $\phi$ is the angle between the star-planet  radius and the line
of  sight.     The positions  were  calculated   at   all times  $t_i$
corresponding  to the  $V_i$ values  of the  light  curve of the star.
When the planet was transiting the star, the light curve was modified,
calculating  the brightness variation   $\Delta V (t_{i})$  and adding
this value to the $V_i$ (see Fig.~\ref{figtransit}).\\

\subsection{Calculating the TDE}

We then selected only the light curves for which there was at least
a  half transit inside  an observing   night  and applied  our transit
detection algorithm. We considered not only  central transits but also
grazing ones.
We considered the number of light curves that exceeded the thresholds,
and also  determined for how many of  these  the transit instants were
correctly identified on the unfolded light curves.

\begin{figure}
   \center
   \includegraphics[width=7.0cm]{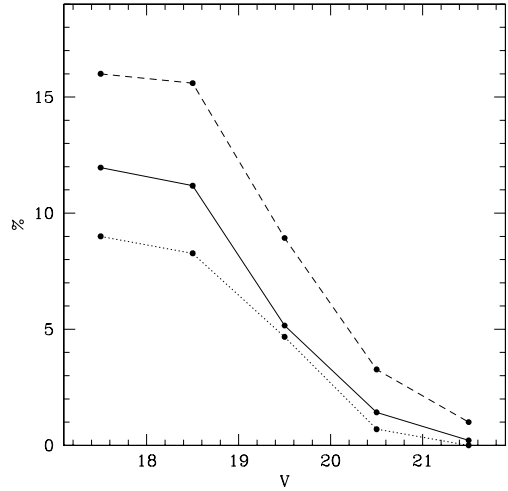}
      \caption{TDE as a function of the stellar magnitude for various assumptions
               on planetary radii distribution.
               From the top to the bottom: 1) Dashed line, $R=(1.4\,\pm\,0.1)\,R_J$;
               2) Solid line, $R=(1.0\,\pm\,0.2)\,R_J$;
               3) Dotted line,  $R=(0.7\,\pm\,0.1)\,R_J$.    
               The adopted threshold for the DSP in this figure is $4.3$.
	       The normalization is respect to the whole number of
               transiting planets.
              }
         \label{fig:effmag}
   \end{figure}

We isolated three different outputs:
\begin{enumerate}

\item Missed candidates: the light curves for which the algorithm
did not get the values of the parameters
that exceeded the thresholds (DSP, OOTV, transit duration and number of
 in transit
points, see Sec~\ref{s:threshold}), or if it did, the epochs of the transits
were not correctly recovered;

\item Partially recovered transit candidates: the parameters exceeded the
 thresholds and at least one
of the transits that fell in the observing window was correctly identified;

\item Totally recovered transit candidates: the parameters exceeded
 the thresholds and all the transits that were
present were correctly recovered.

\end{enumerate}
The TDE was calculated as the sum of the totally and partially
recovered transit candidates relative to the whole number of stars
with transiting planets.
We derive the TDE as a function of magnitude in Fig.~\ref{fig:effmag}.
The   TDE decreases   with  increasing magnitude   because  the  lower
photometric  precision at fainter  magnitudes is not fully compensated
by the  larger transit depth.   The  TDE depends strongly  also on the
assumptions  concerning the planetary  radii,  and on the inclusion of
the  limb darkening effect.  Fig.~\ref{fig:effmag}   is relative to  a
threshold equal to $4.3$ for the DSP (cf.\ Fig.~\ref{fig:effcost}).

The resulting TDE is  about $11.5\%$ around  $V = 18$ and $1\%$ around
$V = 21$ for the case with $R=(1.0\,\pm\,0.2)R_J$.

\begin{figure}
   \center
   \includegraphics[width=9cm]{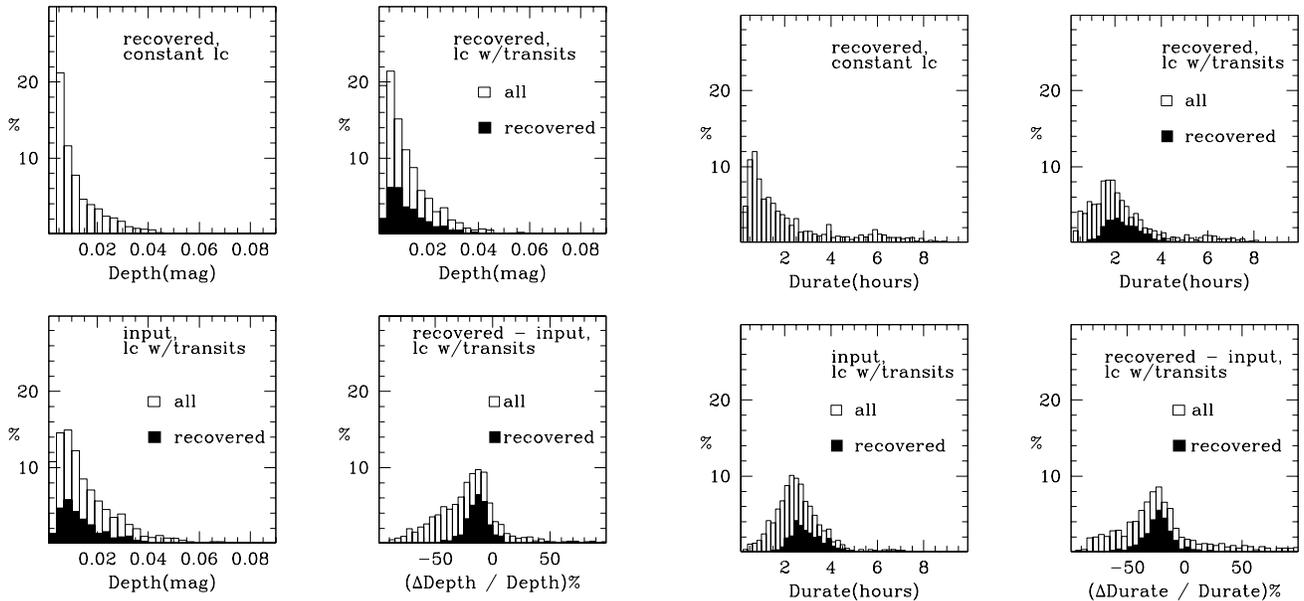}
      \caption{
              (Upper left) Distributions of transit depths measured by the BLS algorithm
	      on the artificial constant-light-curves (lc); (upper right) transit
	      depths measured on artificial light curves with transits
	      added;(lower left) input transit depths used to generate
	      artificial light curves with transits;(lower right) relative
	      difference between the transit depth recovered by BLS
	      and its input value. Empty histograms refer
	      to distributions relative to all light curves,
	      filled ones to light curves with totally and
	      partially recovered transits. Histograms are normalized
	      to all light curves with transiting planets, or,
	      for the upper left panel to all constant light curves.
	      This Figure is relative to CFHT data, and the assumed
	      planetary radii distribution is $R=(1.0\,\pm\,0.2)$.
              }
         \label{fig:depth}
   \end{figure}

\begin{figure}
   \center
   \includegraphics[width=9cm]{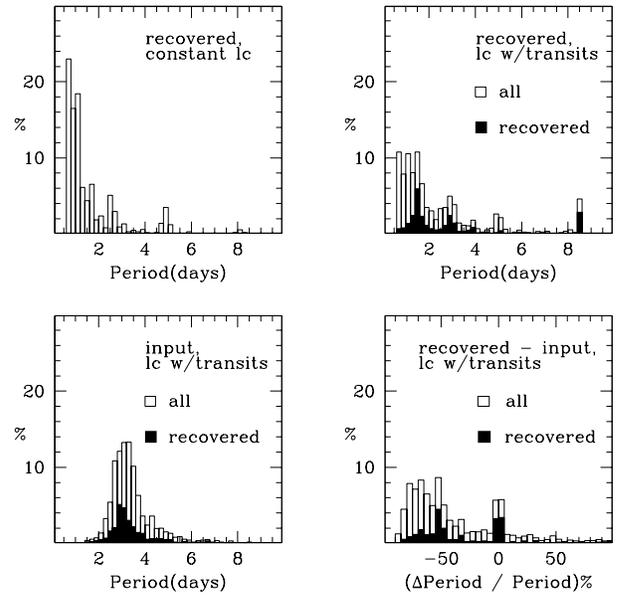}
      \caption{The same as Fig.\ref{fig:depth} for the transit periods.
              }
         \label{fig:period}
   \end{figure}

\begin{figure}
   \center
   \includegraphics[width=9cm]{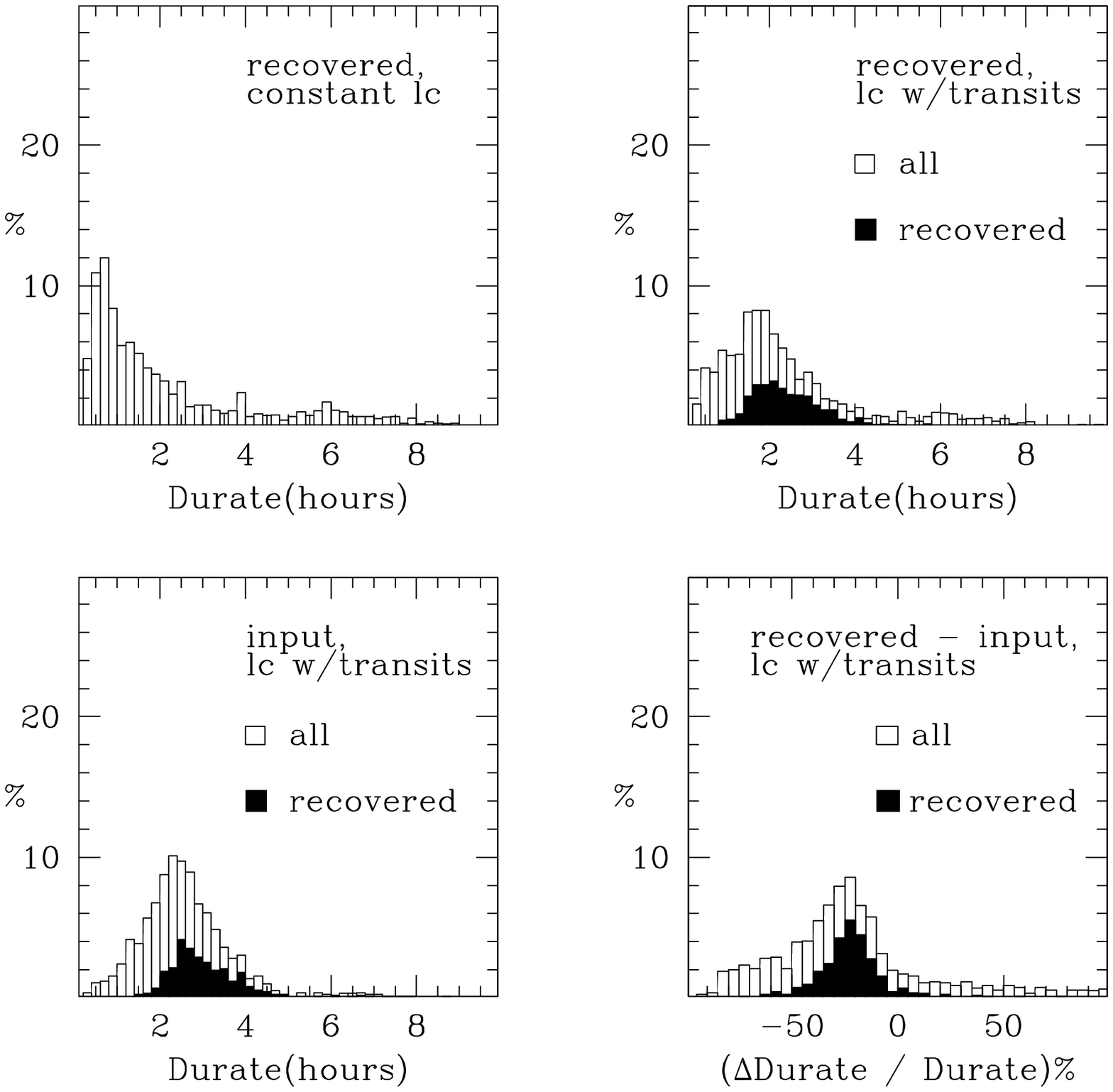}
      \caption{The same as Fig.\ref{fig:depth} for the transit durations.
              }
         \label{fig:durate}
   \end{figure}

Figure~\ref{fig:depth}--\ref{fig:durate}  show the histograms relative
to the input  transit parameters and the recovered  values  of the BLS
algorithm normalized  to the whole number of  transiting planets.  For
comparison we  also show in the  upper  left panel of each  figure the
recovered values  of the BLS  for the constant simulated  light curves
(normalized to the  total number of constant  light curves).  We found
that on  average the BLS  algorithm  has underestimated  the depth and
duration of the transit by about 15-20\%.
This is likely due to the deviation of the transit curves from the box
shape   assumed     by     the   algorithm.   For       the   periods,
(Fig.~\ref{fig:period}), the  recovered    transit period distribution
shown in the upper right panel of Fig.~\ref{fig:period}, had two clear
peaks at $1.5$ and  $3$  days, with the  first  one much more  evident
meaning that the algorithm tends to estimate half of the input transit
period, as shown in the lower panels of the same Figure.
The constant light curve period distribution of  the upper left panel,
instead,  showed   that the   vast  majority of  constant   stars were
recovered with  periods between $0.5$ and  $1$ day, but residual peaks
at $2.5$ and $5$ days were present.

\section{Different approaches in the transit search}
\label{s:diffapproach}
The data   we have  acquired    on \object{NGC~6791} came   from  four
different  sites and involved telescopes  with different diameters and
instrumentations. Moreover, the   observing  window of  each  site was
clearly different with  respect to  the others  as well  as  observing
conditions like seeing, exposure times, etc.

The  first  approach  we tried   consisted   in putting  together  the
observations   coming  from all  the different   telescopes.  The most
important complication we had to face  regarded the different field of
views of the detectors.  This had the consequence that some stars were
measured only in a subset of the sites, and  therefore these stars had
in general different observing windows.
Considering only the   stars  in common would   reduce the  number  of
candidates from $3311$   to $1093$ which   means a reduction of  about
$60\%$ of the targets.
We decided to distinguish eight different    cases, which are shown in
Tab.~\ref{tab:sottocasi}. In the first column a simple binary notation
identifies the different sites: each  digit represents one site in the
following  order: CFHT, SPM, Loiano,  NOT(V) and NOT(I). If the number
correspondent to a generic site  is $1$, it  indicates that the  stars
contained in that     case have been  observed, otherwise the value is
set to $0$. For example,  the notation $11111$  was used for the stars
in common to all 4 sites.
The notation $10000$ indicates the number  of stars which were present
only  on the CFHT  field, and so on.   Each  one of these    cases was
treated  as independent, and the resulting  FAR and expected number of
transiting planets were  added together in order  to  obtain the final
values.

The second approach we followed was to consider only the CFHT data. As
demonstrated in Section~\ref{s:reduction},   overall we  obtained  the
best photometric precision for this data-set.
We considered the  $3311$ candidates which were  recovered in the CFHT
data-set.

For the CFHT data-set,  as shown in Tab.~\ref{tab:sottocasi}, the  DSP
value correspondent to a FAR$=0.1\%$ is equal to $4.3$,
 lower than the other
cases reported in that Table.
Thus, despite the reduced  observing  window of  the CFHT data,  it is
possible to  take advantage of  its increased photometric precision in
the search for planets.

In  Section~\ref{s:candidates},  and in  Section~\ref{s:expected},  we
presented  the candidates and     the different  expected   number  of
transiting planets for these two different approaches.

\begin{table}
\caption{
The   different      cases in which       the    data-sets analysis was
splitted  into.  The  notation in the  first column is explained in the
text, the second column shows the number of  stars in each    case and
the third column refers to the DSP  values assumed, correspondent to a
FAR$=0.1\%$.
\label{tab:sottocasi}
}
\begin{center}
\begin{tabular}{c c c}
\hline
 Case & N.stars & DSP threshold\\
\hline
$11111$ &   $1093$ & 7.5\\
$10000$ &   $771$ & 4.3\\
$10100$ &   $870$ & 5.5\\
$11011$ &   $162$ & 7.1\\
$10001$ &   $112$ & 7.1\\
$11001$ &   $108$ & 7.5\\
$10111$ &   $99$ & 7.2\\
$10011$ &   $96$ & 6.5\\
\hline
\end{tabular}
\end{center}
\end{table}

\section{Presentation of the candidates}
\label{s:candidates}
Table~\ref{tab:candidates} shows the characteristics of the candidates
found    by   the algorithm,   distinguishing   those  coming from the
   entire   data-set   analysis   from  those  coming  from the   CFHT
analysis.

\begin{table*}
\caption{
The  candidates     found  in   the      two cases     discussed    in
Sec.~\ref{s:diffapproach}. The case of the whole data-sets put together
is  indicated with  $ALL$ ($1^{st}$  column), that one  for the $only$
CFHT data-set  is indicated with $CFHT$  ($2^{nd}$ column). A cross (x)
indicates that the candidate was found in that case,  a trait (-) that
it is absent. In the $3^{rd}$ column, the $ID$ of the stars taken from
S03 is shown. Follow the $V$ calibrated  magnitude, the $(B-V)$ color,
the right ascension, ($\alpha$),  and the declination,  ($\delta$), of
the stars.
\label{tab:candidates}
}
\begin{center}
\begin{tabular}{c c c c c c c}
\hline
$ALL$ & $CFHT$ & $ID(Stetson)$ & $V$ & $(B-V)$ & $\alpha(2000)$ & $\delta(2000)$ \\
\hline
 x & - & $6598$ & $18.176$ & $0.921$ & $19^{h}$ $20^{m}$ $48^{s}.65$ & $+37^{\circ}$ $47^{'}$ $41.^{''}1$ \\
 x & - & $4304$ & $17.795$ & $0.874$ & $19^{h}$ $20^{m}$ $41^{s}.39$ & $+37^{\circ}$ $43^{'}$ $28.^{''}9$ \\
 x & - & $4699$ & $17.955$ & $0.846$ & $19^{h}$ $20^{m}$ $42^{s}.67$ & $+37^{\circ}$ $43^{'}$ $31.^{''}5$ \\
 x & x & $1239$ & $19.241$ & $1.058$ & $19^{h}$ $20^{m}$ $25^{s}.42$ & $+37^{\circ}$ $47^{'}$ $45.^{''}2$ \\
 - & x & $4300$ & $18.665$ & $0.697$ & $19^{h}$ $20^{m}$ $41^{s}.38$ & $+37^{\circ}$ $45^{'}$ $23.^{''}3$ \\
 - & x & $7591$ & $18.553$ & $0.959$ & $19^{h}$ $20^{m}$ $51^{s}.51$ & $+37^{\circ}$ $48^{'}$ $58.^{''}7$ \\
\hline
\end{tabular}
\end{center}
\end{table*}

\subsection{Candidates from the whole data-sets}
\label{s:candidateswhole}
\begin{figure}
   \center
   \includegraphics[width=7.5cm]{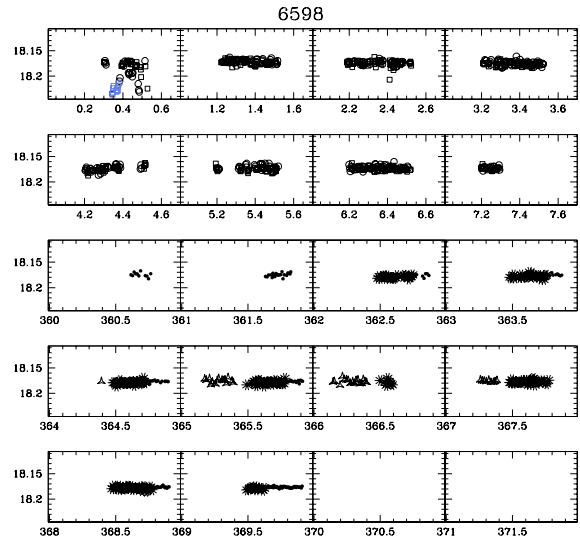}
      \caption{Composite light curve of candidate $6598$. In ordinate
	is reported the calibrated V magnitude and in abscissa the observing epoch,
        (in days), where $0$ corresponds to $JD=52099$.
	{\it Filled circles} indicate CFHT data, {\it crosses} SPM data,
        {\it open triangles} Loiano data, {\it open circles} NOT
        data in the V filter and {\it open squares} NOT data in the
        I filter. Light blue symbols highlight regions which were
	flagged by the BLS.
        }
         \label{fig:6598}
   \end{figure}

\begin{figure}
   \center
   \includegraphics[width=7.5cm]{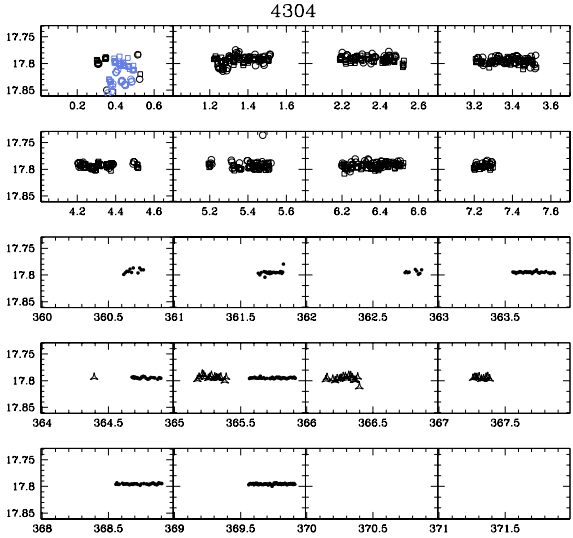}
      \caption{Composite light curve of candidate $4304$.
        }
         \label{fig:4304}
   \end{figure}

\begin{figure}
   \center
   \includegraphics[width=7.5cm]{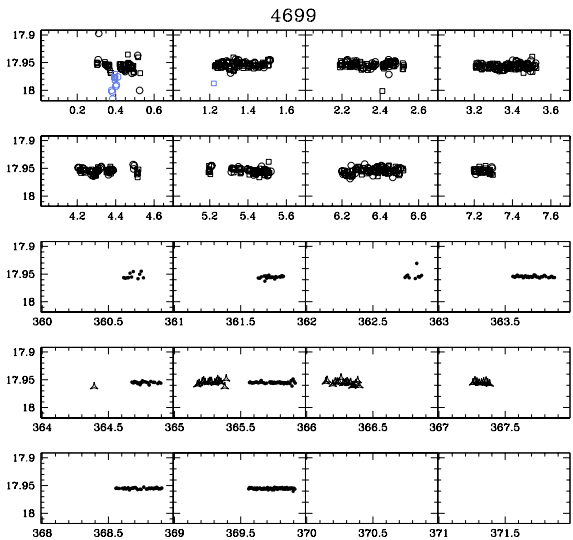}
      \caption{Composite light curve of candidate $4699$.
        }
         \label{fig:4699}
   \end{figure}

Applying    the     algorithm with  the     DSP  thresholds   shown in
Tab.~\ref{tab:sottocasi}  on  the real light  curves  we obtained four
candidates.   Hereafter   we  adopt   the S03   notation   reported in
Tab.~\ref{tab:candidates}.   For  what    concerns  candidates $6598$,
$4304$,                           and                           $4699$
(Fig.~\ref{fig:6598},~\ref{fig:4304},~\ref{fig:4699}) we  noted   (see
also Sec.~\ref{s:screening})  that  the  points contributing    to the
detected  signal   came from the first   observing  night at  the NOT,
meaning  that  bad weather  conditions deeply  affected the photometry
during that night.
In  particular  candidate $6598$, was also   found in the  B03 transit
search survey,    (see Sec.~\ref{s:comparison}),  and flagged     as a
probable spurious candidate. In none of  the other observing nights we
were able  to confirm the  photometric variations which are visible in
the first night  at the NOT.  We concluded that these three candidates
are of spurious nature.

\begin{figure}
   \center
   \includegraphics[width=7.5cm]{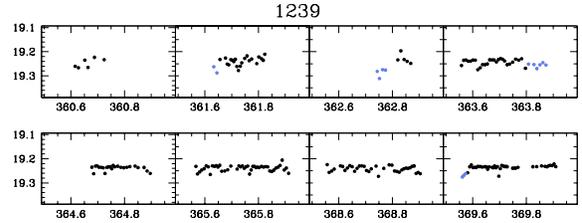}
      \caption{CFHT light curve for candidate $1239$.
        }
         \label{fig:1239}
   \end{figure}

The fourth candidate corresponds to   star $1239$, that is located  in
the external regions  of the cluster. For this  reason we presented in
Fig.~\ref{fig:1239} only the data coming from the  CFHT. In this case,
the data points  appear irregularly scattered underlying  a particular
pattern of variability or simply a spurious photometric effect.

\subsection{Candidates from the CFHT data-set}

Considering only  the  data coming  from   the CFHT observing run   we
obtained three candidates. The star $1239$ is  in common with the list
of candidates coming from the   whole data-sets because, as  explained
above, it is located in the external regions for which we had only the
CFHT data.
For  candidate   $4300$,  the    algorithm  identified   two    slight
($\sim\,0.004\,$ mag) magnitude variations with duration of around one
hour during the  sixth and the tenth night,  with  a period of  around
$4.1$ days.  A jovian planet around  a main sequence star of magnitude
V=$18.665$, (with  $R=0.9\,R_{\odot}$, see Fig.~\ref{fig_vmr}), should
determine a  transit with  a  maximum  depth  of around  $1.2\%$,  and
maximum duration of $2.6$ hours.
Although compatible with  a grazing transit, we  observed that the two
suspected eclipses are not identical,  and, in any case, outside these
regions, the photometry appears quite scattered. Star $7591$, instead,
does not show any significant feature.

From  the   analysis of  these  candidates   we concluded that
 no transit features are detected for both the entire data-sets
and the CFHT   data. Moreover,  we can say   to have   recovered  the
expected number of  false alarm candidates which was $(3.3\,\pm\,1.3)$
as explained in Sec.~\ref{s:whole}.

   \begin{figure}
   \center
   \includegraphics[width=7.5cm]{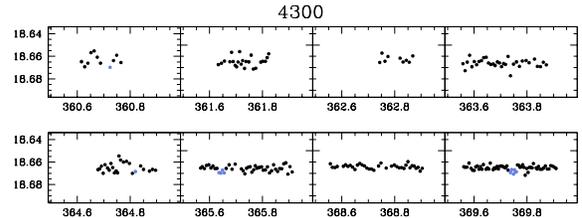}
      \caption{CFHT light curve for candidate $4300$.
        }
         \label{fig:4300}
   \end{figure}

   \begin{figure}
   \center
   \includegraphics[width=7.5cm]{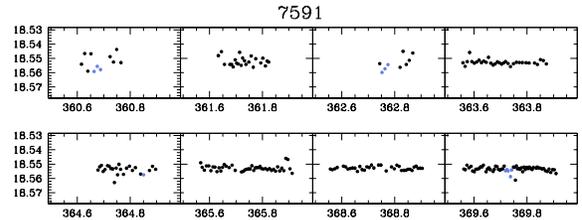}
      \caption{CFHT light curve for candidate $7591$.
        }
         \label{fig:7591}
   \end{figure}

\section{Expected number of transiting planets}
\label{s:expected}

\subsection{Expected frequency of close-in planets in \object{NGC~6791}}
\label{s:planetfreq}

The  frequency    of  short-period planets   in~\object{NGC~6791}  was
estimated considering the enhanced  occurrence of giant planets around
metal  rich  stars  and the  fraction   of  hot  Jupiters  among known
extrasolar planets.

Fischer \& Valenti (\cite{fischer05}) derived the probability $\mathcal{P}$ of
formation of  giant planets with orbital period shorter than 4$\,$yr and
radial velocity semi-amplitude $K>30$ ms$^{-1}$ as a function of [Fe/H]:
\begin{equation}\label{fv04}
\mathcal{P}=0.03 \cdot 10^{\,2.0\,{\rm [Fe/H]}} \hspace{1cm} -0.5<{\rm [Fe/H]}<0.5
\end{equation}

The  number of stars with  a giant planet  with  $P<9$ d was estimated
considering the ratio  between the number  of  the planets with  $P<9$
days  and  the total number  of  planets from  Table   3 of Fischer \&
Valenti~\cite{fischer05}  (850       stars  with     uniform    planet
detectability). The result is $0.22^{+0.12}_{-0.09}$.

Assuming for~\object{NGC~6791} [Fe/H]$=+0.17$ dex, a conservative
lower limit to the cluster metallicity, from Equation
\ref{fv04} we determined that the probability that a
cluster star   has  a giant   planet with  $P<9$  is 1.4\%.   Assuming
[Fe/H]$=+0.47$, the  metallicity resulting from the  spectral analysis
by Gratton et al.~(\cite{gratton06}), the probability rises to 5.7\%.

Our estimate assumes that the planet period and the metallicity of the
parent star  are  independent,   as  found  by  Fischer   \&   Valenti
(\cite{fischer05}).  If the hosts of  hot Jupiters are even more metal
rich than  the hosts of  planets with longer  periods, as  proposed by
Society (\cite{sozzetti04}),  then the expected frequency of close-in
planets at the  metallicity  of \object{NGC~6791} should  be  slightly
higher than our estimate.

\subsection{Expected number of transiting planets}
\label{s:expectedtr}

In order to evaluate the expected number of transiting planets in our survey we followed this
procedure:
\begin{itemize}
\item From the constant stars of our simulations (see Sec.~\ref{s:simul}),
 taking into account the luminosity function of main sequence stars of the
 cluster, we randomly selected a sample corresponding to the probability
 that a star has a planet with $\rm P\leq9$ d.
\item From the $V$ magnitude of the star we calculated the mass and radius.
\item To each star in this sample we assigned a planet with mass,
  radius, period randomly chosen from the distributions described
  in Sec.~\ref{s:planetpar}, and $cos\,i$ randomly chosen inside the
  range $0\,$-$\,1$.
 The range spanned for the periods was
  $\rm 1\,<\,P\,<\,9$ days, with a step size of $0.006$ days. For
  planetary radii we considered the three distributions described in
  Sec.~\ref{s:planetpar}, sampled with a step size of $\rm
  0.001\,R_J$, and inclinations were varied of $0.005$ degrees.

\item We selected {\em only} the stars with planets that can make transits thanks to their
      inclination angle given by the relation:
\[
\rm \cos i \leq \frac{R_{pl}+R_{\star}}{a}
\]
\item Finally, as described above, we assigned to each planet the
      initial phase $\phi_0$ and the revolution
      orbital direction $s$ and modified the constant light curves inserting the transits.
      The initial phase was chosen randomly inside the range
      $0$-$360$ degrees, with a step size of $0.3$ degrees.

\item We applied the BLS algorithm to the modified light curves with
  the adopted thresholds.

We performed $7000$ different simulations and we calculated the mean
values of these quantities:

\begin{itemize}
\item The number of MS stars with a planet:  $N_{pl}$
\item The number of planets that make transits (thanks to their inclination angles): $N_{geom}$
\item The number of planets that make {\em one or more} transits in the observing window: $N_{+1}$
\item The number of planets that make {\em one} single transit in the observing window: $N_{1}$
\item The number of transiting planets detected by the algorithm
for the three different planetary radii distributions adopted,
(as described in Sect.~\ref{s:simul}), $R^{1}=(0.7\,\pm\,0.1)\,R_J$,
$R^{2}=(1.0\,\pm\,0.2)\,R_J$, and $R^{3}=(1.4\,\pm\,0.1)\,R_J$.
\end{itemize}
\end{itemize}

\subsection{FAR and expected number of detectable transiting planets
  for the whole data-sets}
\label{s:whole}

We followed  the procedure reported  in Sec.~\ref{s:simul}  to perform
simulations with the artificial  stars. It is  important to  note that
artificial  stars  were added  exactly  in the   same positions in the
fields of  the  different  detectors.   This is  important  because it
assured the  homogeneity of the artificial  star tests.  We decided to
accept a FAR equal to $0.1\%$, which meant that  we expected to obtain
$(3.3\,\pm\,1.3)$ false alarms from the total number of $3311$ cluster
candidates.  The DSP  thresholds  correspondent to this FAR  value are
different for each case, and is reported in Table~\ref{tab:sottocasi}.

Table~\ref{t:results_sim_total}   displays  the     results for    the
simulations  performed  in order  to  obtain  the expected  numbers of
detectable  transiting   candidates for three    values of [Fe/H] (the
values  found by    Carraro et  al.~\cite{carraro06} and Gratton    et
al.~\cite{gratton06} and a   conservative lower limit  to the  cluster
metallicity).

The  columns listed  as    $N_{geom}$, $N_{1+}$ and $N_{1}$   indicate
respectively the number of   planets which have a  favorable geometric
inclination  for  the  transit,  the number of  expected  planets that
transit at  least one time within the  observing window and the number
of  expected planets  that transit  exactly one time  in the observing
window.

The numbers of  expected  transiting planets in  our  observing window
detectable  by the algorithm were  calculated  for the three different
planetary  radii  distributions (see Sec.~\ref{s:simul},  and previous
paragraph).  On the  basis of the current  knowledge  on giant planets
the most likely case corresponds to $R^{2}=(1.0\,\pm\,0.2)\,R_J$.

Table~\ref{t:results_sim_total} shows that,  assuming the most  likely
planetary radii  distribution  $R=(1.0\,\pm\,0.2)\,R_J$ and  the high
metallicity resulting from recent  high dispersion studies (Carraro et
al.~\cite{carraro06}; Gratton et al.~\cite{gratton06}), we expected to
be  able to detect $2-3$ planets  that exhibit at least one detectable
transit in our observing window.

\begin{table*}
\caption{The Table shows the results of our simulations on the expected number of
 detectable transiting planets for the whole data-set (all the cases of
 Tab.~\ref{tab:sottocasi}) as explained in
 Sect.~\ref{s:whole}. $N_{geom}$ indicates planets with favorable
 inclination for transits, $N_{1+}$, and $N_{1}$,
 planets that transit respectively at least one time
 and only one time inside the observing window. $R^{1}$, $R^{2}$,
 $R^{3}$, indicate the expected number of detectable transiting planets inside
our observing window, for the three assumed planetary radii
 distributions, (see Sec.~\ref{s:planetpar}).
}
\label{t:results_sim_total}
\centering
\begin{tabular}{c c c c c c c}
\hline
  [Fe/H]  & $N_{geom}$ & $N_{1+}$ &  $N_{1}$ & $R^{1}$ &  $R^{2}$ & $R^{3}$ \\
\hline
  +0.17   &  5.39  &  3.08  & 1.68   &  {$\bf 0.0\,\pm\,0.0$} &  {$\bf 0.0\,\pm\,0.0$}  &  {$\bf 1.8\,\pm\,0.9$} \\
  +0.39   & 15.13  &  8.32  & 4.60   &  {$\bf 0.1\,\pm\,0.1$} &  {$\bf 1.9\,\pm\,0.8$}  &  {$\bf 3.6\,\pm\,1.8$} \\
  +0.47   & 21.92  & 11.95  & 6.62   &  {$\bf 0.2\,\pm\,0.3$} &  {$\bf 3.2\,\pm\,1.9$}  &  {$\bf 5.4\,\pm\,1.8$} \\
\hline
\end{tabular}
\end{table*}

\subsection{FAR and expected number of detectable transiting planets for the CFHT data-set}
Table~\ref{t:results_sim_cfh} shows  the expected number of detectable
planets in our observing  window for the case  of  the $CFHT$ data.  A
comparison      with Table~\ref{t:results_sim_total} revealed that, in
general, except for the largest planetary radii distribution, $R^{3}$,
the number of expected  detections  is not increasing considering  all
the sites together instead of the  $CFH$ only. 
Moreover,  for  the  cases   of   [Fe/H]$=(+0.39,+0.47)$dex, and   the
$R=(0.7\,\pm\,0.1)R_{J}$ radii distribution, we obtained significantly
better results considering only  the $CFHT$ data than putting together
all the data-sets. We interpreted this result as  the evidence that the
transit signal is,  in general, lower  than  the total scatter in  the
composite light  curves and this   didn't allow the algorithm to  take
advantage of  the increased observing window giving,  for the cases of
major   interest, $R=(1\,\pm\,0.2)R_J$ and  [Fe/H]$=(+0.39,+0.47)$dex,
comparable results.

\begin{table*}
\caption{The same as \ref{t:results_sim_total}, but for the case of
 the only $CFHT$ data as explained in Sect.~\ref{s:expectedtr}.}
\label{t:results_sim_cfh}
\centering
\begin{tabular}{c c c c c c c}
\hline
  [Fe/H]  & $N_{geom}$ & $N_{1+}$ &  $N_{1}$ & $R^{1}$ &  $R^{2}$ & $R^{3}$ \\
\hline
  +0.17   &  5.39  &  2.49  &   1.98  &  {$\bf 0.2\,\pm\,0.5$} &  {$\bf 0.4\,\pm\,0.7$}  &  {$\bf 0.6\,\pm\,0.8$} \\
  +0.39   & 15.13  &  7.01  &   5.39  &  {$\bf 1.6\,\pm\,1.3$} &  {$\bf 2.3\,\pm\,1.6$}  &  {$\bf 2.6\,\pm\,1.7$} \\
  +0.47   & 21.92  & 10.12  &   7.94  &  {$\bf 2.5\,\pm\,1.7$} &  {$\bf 3.4\,\pm\,2.0$}  &  {$\bf 4.0\,\pm\,2.1$} \\
\hline
\end{tabular}
\end{table*}

\section{Significance of the results}
\label{s:significance}
As explained in Sec.~\ref{s:candidates},  on real data we obtained $4$
candidates, considering the data coming  from the entire data-sets, (all
the     cases  of  Tab.~\ref{tab:sottocasi}),   and   $3$   candidates
considering only the  best photometry coming from  the $CFHT$. None of
these candidates shows clear transit features, and their number agrees
with the   expected number of    false   candidates coming  from   the
simulations $(3.3\,\pm\,1.3)$ as explained in Sec.~\ref{s:far}.

Considering  the  case  relative   to the  metallicity  of  Carraro et
al.~\cite{carraro06} ([Fe/H]$=+0.39$) and  the  one  relative  to  the
metallicity of  Gratton et al.~\cite{gratton06}, ([Fe/H]$=+0.47$), and
given the   most   probable    planetary  radii   distribution    with
$R=(1.0\,\pm\,0.2)R_J$,    from   Table~\ref{t:results_sim_total}  and
Table~\ref{t:results_sim_cfh} we expected between  $2$ and $3$ planets
with at least one detectable transit inside our observing window.

Therefore, this study reveals a lack of transit detections.

What is  the  probability that  our survey  resulted  in no transiting
planets just by chance? To  answer this question we  went back to  the
simulations described  in Sect.~\ref{s:expectedtr} and  calculated the
ratio of the  number  of simulations for  which we  were not  able  to
detect  any  planet relative    to the   total number   of simulations
performed.
The  resulting probabilities to    obtain no transiting  planets  were
respectively around $10\%$ and $3\%$ for  the metallicities of Carraro
et al.~\cite{carraro06} and Gratton et al.~\cite{gratton06} considered
above.

\section{Implication of the results}
\label{s:implications}

Beside the rather  small, but not negligible  probability of  a chance
result,    ($3$-$10$\%,   see  Sec.~\ref{s:significance}),   different
hypothesis can be invoked to explain the lack of observed transits. We
have discussed them here.

\subsection{Lower frequency of close-in planets in cluster environments}
The lack of  observed transits might  be due  to a lower  frequency of
close-in planets  in clusters compared  to the field  stars of similar
metallicity.  In general, two   possible factors could prevent  planet
formation especially in clustered environments:

\begin{itemize}
\item {in the first million years of the cluster life, UV-flux
 can evaporate fragile embryonic dust
disks from which planets are expected to form.
Circumstellar disks associated with solar-type stars can be readily
evaporated in sufficiently large clusters, whereas disks around
smaller (M-type) stars can be evaporated in more common, smaller groups.
 In addition, even though giant planets could still form in
 the disk region $r\,=\,5$-$15$ AU, little disk mass (outside that
 region) would be available to drive planet migration.;}
\item {on the other hand, gravitational forces could
 strip nascent planets from their parent
stars or, taking in mind that transit planet searches are biased toward
'hot jupiter' planets, tidal effects could prevent the planetary migration
processes which are essential for the formation of this kind of
planets.}
\end{itemize}

These factors  depend  critically  on   the cluster size.     Adams et
al.~(2006), show that for   clusters with $100$-$1000$ members  modest
effects are  expected on forming  planetary  systems.  The interaction
rates are low, so that  the typical solar  system experiences a single
encounter with  closest approach distance  of $1000$ AU. The radiation
exposure is also low, so  that photo-evaporation of circumstellar disks
is only    important beyond   $30$  AU.   For  more  massive  clusters
like~\object{NGC6791}, these factors are  expected to  be increasingly
important  and could drastically  affect planetary formation (Adams et
al. 2004).

\subsection{Smaller planetary radii for planets around very metal rich host stars}

Guillot et al.~(2006) suggested that  the masses of heavy elements  in
planets   was  proportional to  the    metallicities of  their  parent
star. This correlation remains to be confirmed, being still consistent
with a no-correlation  hypothesis at  the  $1/3$  level in the   least
favorable case.  A  consequence of this  would be a smaller radius for
close-in planets orbiting super-metal rich stars.
Since the transit  depth scales with  the square  of the radius,  this
would   have    important    implications for   ground-based   transit
detectability,                                                    (see
Tables~\ref{t:results_sim_cfh}-~\ref{t:results_sim_total}).

\subsection{Limitations on the assumed hypothesis}
While we exploited the best  available results to estimate the expected
number of   transiting   planets, it  is possible  that   some  of our
assumptions   are  not  completely realistic,  or    applicable to our
sample. One  possibility  is that  the  planetary frequency no  longer
increases above a given metallicity.  The small number of stars in the
high  metallicity  range in  the Fischer \&  Valenti  sample makes the
estimate of  the  expected planetary  frequency for  the most metallic
stars   quite  uncertain.     Furthermore,  the  consistency  of   the
metallicity scales of Fischer \& Valenti (2005), Carraro et al. (2006)
and Gratton et al. (2006) should be checked.

Another  possibility   concerns  systematic   differences between  the
stellar sample studied  by Fischer  \&  Valenti, and the present  one.
One relevant  point is  represented by  binary systems. The  sample of
Fischer \& Valenti  has some  biases  against binaries, in  particular
close binaries. As the frequency of  planets in close binaries appears
to be   lower than that   of planets orbiting  single  stars  and wide
binaries (Bonavita \& Desidera  2007, A\&A, submitted), the  frequency
of planets in the Fischer \& Valenti sample should be larger than that
resulting in an  unbiased sample. On the other  hand, our selection of
cluster stars  excludes the stars   in the binary  sequence, partially
compensating this effect.

Another  possible  effect  is   that of   stellar  mass.  As  shown in
Fig.~\ref{fig_vmr},   the cluster's stars  searched  for transits have
mass between $1.1$ to $0.5$ M$_\odot$. On the other hand, the stars in
the FV  sample have  masses between  $1.6$ to $0.8$  M$_\odot$. If the
frequency  of  giant planets depends on  stellar  mass, the results by
Fischer \&  Valenti (2005) might   not be directly  applicable to  our
sample.

Furthermore,  some  non-member contamination is certainly  present. As
discussed in Section~\ref{s:selection},   the  selection   of  cluster
members was done photometrically around a fiducial main sequence line.

\subsection{Possibility of a null result being due to chance}

As shown in Sec.~\ref{s:significance}, the probability that our null result
was simply due to chance was comprised between $3$\% and $10$\%, depending
on the metallicity assumed for the cluster. This is a rather small, but not negligible
probability, and other efforts must be undertaken to reach a firmer
conclusion.

\section{Comparison of the transit search surveys on \object{NGC~6791}}
\label{s:comparison}

It is important to compare our results on the presence of planets with
those of other  photometric   campaigns performed in past    years. We
consider in this comparison B03 and M05.

\subsection{The Nordic Optical Telescope (NOT) transit search}

As already described in this paper, (e.g. see Sect.~\ref{s:observations}),
 in July $2001$, at NOT, B03  undertook a transit search on
\object{NGC~6791} that lasted eight nights. Only seven of these nights were good enough
to   search   for planetary  transits.  Their time   coverage was thus
comparable to the CFHT data presented here.
The expected number of transits was obtained considering as candidates
all  the stars with photometric precision  lower than $2\%$, (they did
not isolate cluster  main  sequence stars, as  we  did, but they  then
multiplied  their  resulting expected numbers  for  a  factor equal to
$85\%$ in  order  to  account  for binarity),  and assuming  that  the
probability that  a local G or   F-type field star  harbors a close-in
giant planet is around $0.7\%$.
With these and other obvious  assumptions B03 expected $0.8$  transits
from their  survey. However,  they made also   the hypothesis that for
metal-rich stars the fraction of  stars harboring planets is  $\sim10$
times  greater than  for   general  field stars,  following   Laughlin
(\cite{laughlin00}).
In this   way, they would  have  expected  to  find ``at  least  a few
candidates  with single  transits''.  In Section~\ref{s:reduction}  we
showed how the  photometric precision for  the  NOT was  in general of
lower quality for the brightest stars with respect  to that one of SPM
and    Loiano.    This     fact  can    be      recognized  also    in
Table~\ref{tab:sottocasi} where the value of the threshold for the DSP
was always bigger than $6.5$ when the NOT observations were included.
This  demonstrates the higher noise level  of this data-set. We did not
perform  the accurate analysis   of the expected number of  transiting
planets  considering only the  NOT  data,  but,  on  the basis  of our
assumptions,  and on the  photometric precision  of  the NOT data, the
numbers showed in Table~\ref{t:results_sim_cfh} for the CFHT should be
considered  as an upper  limit for the expected  transit  from the NOT
survey.

B03 reported  ten transit events, two  of which, (identified in B03 as
T6 and T10),  showed double  transit  features,  and the others   were
single transits.  Except for candidate T2, which was recovered also in
our analysis (see Sec.~\ref{s:candidateswhole})  our algorithm did not
identify any other of the candidates reported by B03.

\begin{figure}
   \center
   \includegraphics[width=7.5cm]{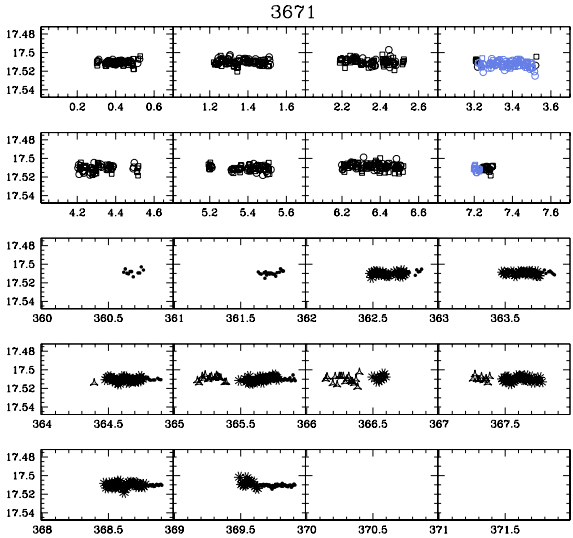}
      \caption{Composite light curve for
	candidate $3671$ correspondent
        to T5 of BO3. Different symbols have
	the same meaning of Fig.~\ref{fig:6598}.
        }
         \label{fig:3671}
   \end{figure}

\begin{figure}
   \center
   \includegraphics[width=7.5cm]{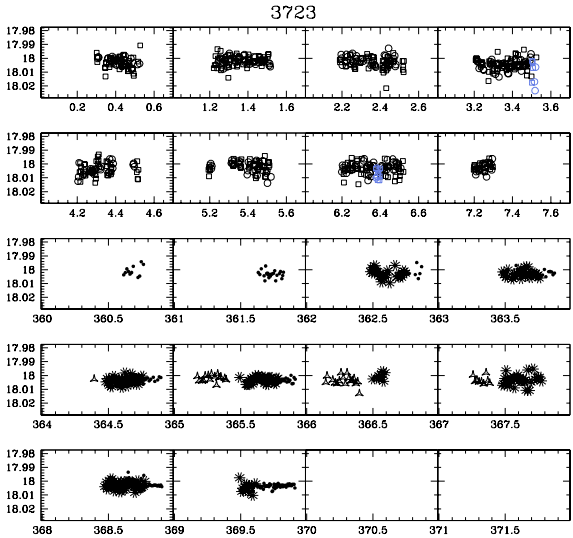}
      \caption{Composite light curve
	for candidate $3723$ correspondent
        to T7 of BO3.
        }
         \label{fig:3723}
   \end{figure}

B03 recognized that most of the candidates were likely spurious, while
three cases, referred  as T5,  T7 and  T8,  were  considered the  most
promising ones.  We noted that T8 lies off the cluster main sequence.
Therefore,  it   can  not  be  considered    as   a  planet  candidate
for~\object{NGC~6791}. Furthermore, from our CFHT images we noted that
this  candidate is likely  a blended star.   The  other two candidates
were on the  main sequence of~\object{NGC~6791}. Visual inspection  of
the light curves  in Fig.~\ref{fig:3671}  and Fig.~\ref{fig:3723} also
show no sign of eclipse.

\begin{figure}
   \center
   \includegraphics[width=7.5cm]{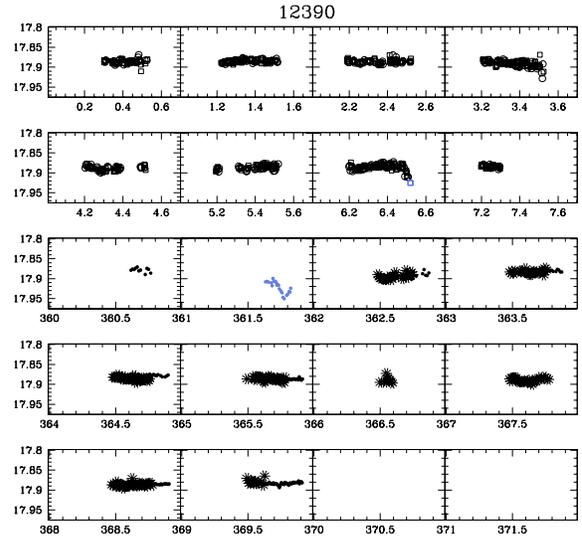}
      \caption{Composite light curve
	for candidate $12390$ correspondent
        to T9 of BO3.
        }
         \label{fig:12390}
   \end{figure}

Finally,  candidate T9, (Fig.~\ref{fig:12390})   lies off the  cluster
main sequence  and  it  was recognized by    B03 to be   a long-period
low-amplitude variable (V80).  In our photometry, it shows clear signs
of variability,  and a $\sim0.05$  mag eclipse during the second night
of the CFHT campaign at  $t=361.8$, and probably  a partial eclipse at
the end of the  seventh night of the  NOT data-set, at  $t=6.4$, ruling
out  the possibility of  a planetary   transit, because the  magnitude
depth of  the  eclipse is much larger  than   what is  expected  for a
planetary transit.\\

It is not surprising that almost all of the candidates reported by B03
were not  confirmed   in  our  work,  even   for the   NOT  photometry
itself. Even if the   photometry  reduction algorithm was   the  same,
(image subtraction, see  Sec.~\ref{s:reduction}), all  the other steps
that followed, and  the selection criteria  of the  candidates were in
general different.  This, in turn, reinforces the  idea  that they are
due to spurious photometric effects.

\subsection{The PISCES group extra-Solar planets search}

The  PISCES  group     collected $84$  nights      of  observations on
\object{NGC~6791}, for a total of $\sim\,300$ hours of data collection
from July $2001$  to July $2003$,  at the $1.2$m Fred Lawrence Whipple
Observatory   (M05).   Starting   from their  $3178$   cluster members
(selected  considering  all the main  sequence   stars with  RMS$ \leq
5\%$), assuming a distribution of  planetary radii between $0.95\,R_J$
and $1.5\,R_J$,  and a  planet frequency of   $4.2\%$, M05 expected to
detect $1.34$ transiting planets  in the cluster. They didn't identify
any transiting candidate.
 Their planet    frequency is   within  the    range that   we assumed
 ($1.4\%$--$5.7\%$).  Our number of   candidate main-sequence stars is
 slightly in excess relative  to that of M05, even  if their  field of
 view  is  larger than   our   own  ($\sim\,23\,$arcmin$^{2}$  against
 $\sim\,19\,$arcmin$^{2}$ of S03  catalog),  since we were able   to
 reach $\sim\,2\,$mag  deeper  with  the  same photometric   precision
 level. Their  number of expected  transiting  planets is  of the same
 order  of magnitude  as  our  own because    of their  huge  temporal
 coverage.   In any case, looking  at  figure $7$  of  M05, one should
 recognize that their detection   efficiency greatly favors  planetary
 radii larger  than $1\,R_J$.    A  more realistic  planetary   radius
 distribution,      for      example   $(1.0\,\pm\,0.2)\,R_J$,  should
 significantly decrease their expectations,  as recognized by the same
 authors.

\section{Future investigations}
\label{s:future}

\object{NGC~6791} has been recognized as one of the most promising targets for studying
the  planet formation  mechanism in   clustered environments, and  for
investigating  the  planet frequency as  a function  of  the host star
metallicity.  Our estimate    of  the expected  number  of  transiting
planets, (about  15--20  assuming the metallicity  recently derived by
means of  high-dispersion  spectroscopy by Carraro   et  al. 2006, and
Gratton et  al. 2006, and  the planet frequency  derived by Fischer \&
Valenti 2005),  confirms  that this is   the best open  cluster  for a
planet search.

However, in spite of fairly ambitious observational efforts 
by different groups,
no firm conclusions  about  the presence or   lack of planets  in  the
cluster can be reached.

With the goal of understanding the  implications of this result and to
try  to  optimize    future   observational  efforts,   we  show,   in
Table~\ref{tab:comparison}, that the number of hours collected on this
cluster with  $>3$ m telescopes is  much lower than the time dedicated
with $1-2$ m class telescopes.
Despite the fact that we were able to get adequate
photometric precisions even with $1-2$ m class telescopes,
(see Sec.~\ref{s:reduction}),
in general smaller aperture telescopes are typically located
on sites with poorer observing conditions, which limits the
temporal sampling and their 
photometry is characterized by larger systematic effects.
As  a result,  the number of  cluster  stars with adequate photometric
precision for planet transit  detections is quite limited.  Our  study
suggests that  more extensive photometry with  wide field imagers at 3
to 4-m  class telescopes (e.g.  CFHT) is required  to reach conclusive
results on the frequency of planets in \object{NGC~6791}.

 We  calculated that, extending  the  observing window to  two transit
 campaigns of ten  days   each, providing  that the  same  photometric
 precision we had  at the CFHT could  be reached, we could reduce  the
 probability of null detection to $0.5$\%.

\begin{table*}[!]
\caption{Number of nights and  hours which have
 been devoted to the study of \object{NGC~6791} as a function of the
 diameter of the telescope used for the survey.
 We adopted a mean of $5$ hours of observations per night.
\label{tab:comparison}
}

\begin{center}
\begin{tabular}{c c c c c}
\hline
Telescope & Diameter(m) & N$_{nights}$ & Hours & Ref.\\
\hline
FLWO &   1.2     &     84     & $\sim$300 & M05\\
Loiano &   1.5     &      4     &      20 & This paper \\
SPM &   2.2     &      8     &      48 &  This paper\\
NOT &   2.54    &      7     &      35 & B03 and this paper\\
CFHT &   3.6     &      8     &      48 &  This paper \\
MMT &   6.5     &      3     &      15 & Hartmann et al.~(2005)\\
\hline

\end{tabular}
\end{center}
\end{table*}

\section{Conclusions}
\label{s:conclusions}

The main purpose of this work was to investigate the problem of planet
formation  in stellar open clusters.  We  focused our attention on the
very  metal   rich open  cluster   \object{NGC~6791}.   The idea  that
inspired this  work  was that looking  at more  metal rich   stars one
should expect a higher  frequency of planets, as  it has been observed
in the solar neighborhood   (Santos et al.  2004, Fisher  \& Valenti,
2005).
Clustered environments  can be  regarded as astrophysical laboratories
in which    to  explore planetary frequency  and   formation processes
starting from a well defined and homogeneous sample  of stars with the
advantage  that   cluster  stars   have  common   age,   distance, and
metallicity.
As shown in  Section~\ref{s:observations}, a huge observational effort
has been  dedicated to  the study  of  our  target cluster  using four
different  ground based telescopes,  (CFHT, SPM, Loiano, and NOT), and
trying to take advantage from multi-site simultaneous observations. In
Section~\ref{s:reduction}, we showed   how   we were able   to  obtain
adequate photometric precisions  for  the transit  search for  all the
different  data-sets (though  in different magnitude  intervals).  From
the detailed simulations described in Section~\ref{s:expected}, it was
demonstrated that, with  our best photometric  sequence,  and with the
most realistic assumption  that the  planetary radii  distribution  is
$\rm R=(1.0\,\pm\,0.2)R_{J}$, the expected number of detectable transiting
planets with   at least one transit  inside  our  observing window was
around $2$, assuming as cluster metallicity [Fe/H]=$+0.39$, and around
$3$ for [Fe/H]$=+0.47$.
Despite  the number  of  expected positive  detections, no significant
transiting planetary   candidates  were found   in our  investigation.
There was a rather small,  though not negligible probability that  our
null  result   can   be simply  due   to   chance,   as explained   in
Sect.~\ref{s:significance}: we   estimated  that this   probability is
$10\%$  for [Fe/H]$=+0.39$, and  $3\%$  for [Fe/H]$=+0.47$.   Possible
interpretations    for    the      lack  of     observed      transits
(Sect.~\ref{s:implications})  are   a  lower frequency   of   close-in
planets around solar-type stars in  cluster environments with  respect
to field stars, smaller planetary radii for planets around super metal
rich stars,   or some limitations in  the  assumptions adopted  in our
simulations.  Future investigations with $3$-$4$m class telescopes are
required (Sect~\ref{s:future})   to   further  constrain the planetary
frequency  in \object{NGC~6791}. Another twenty  nights with this kind
of instrumentation are necessary to   reach a firm conclusion on  this
problem.  The uniqueness  of \object{NGC~6791}, 
which is the only galactic open cluster for which  we expect more than
10  giant planets   transiting  main  sequence stars    if  the planet
frequency is the same as for field stars of similar metallicity, makes
such  an    effort  crucial  for exploring the     effects of  cluster
environment on planet formation.

\newpage

\begin{acknowledgements}
We warmly thank M.~Bellazzini and F.~Fusi Pecci for having made possible
the run at Loiano Observatory. \\
This work was partially funded by COFIN 2004
 ``From stars to planets: accretion, disk evolution and
planet formation'' by Ministero Universit\'a e Ricerca
Scientifica Italy.\\
 We thanks the referee, Dr. Mochejska, for
useful comments and suggestions allowing the improvement of the paper.
\end{acknowledgements}

{}

\end{document}